\documentclass[11pt,a4paper]{article}
\usepackage{jcappub}
\usepackage{epsfig}   
\usepackage{graphics}

\newcommand{\om}{\Omega_{\rm M}}
\newcommand{\ok}{\Omega_{\rm K}}
\newcommand{\ola}{\Omega_{\Lambda}}

\title{A model independent measure of the large scale curvature of the Universe}

\author[a]{Edvard M\"ortsell,} 
\author[a,b]{Jakob J\"onsson}

\affiliation[a]{The Oskar Klein Centre for Cosmoparticle Physics, 
        Department of Physics, \\ Stockholm University,  AlbaNova
        University Center, S-106 91 Stockholm, Sweden}
\affiliation[b]{Department of Physics, Oxford University, \\Denys Wilkinson Building, 
	Keble Road, Oxford, OX1 3RH, UK}
         
\emailAdd{edvard@fysik.su.se}
\emailAdd{jacke@fysik.su.se}

\dedicated{Dedicated to Lasse O'M\aa nsson, the late editor-in-chief of 
Svenska Mad.}

\abstract{
Cosmological distances as a function of redshift depend on the
effective curvature density, $\ok$, via the effect on the geometrical
path of photons from large scale spatial curvature and its effect on
the expansion history, $H(z)$. Cosmological time, however, depends on
the expansion history only. Therefore, by combining distance and
lookback time observations (or other estimates of the expansion
history), it is possible to isolate the geometrical curvature
contribution and measure $\ok$ in a model independent way, i.e.,~free
from assumptions about the energy content of the universe.

We investigate two different approaches to accomplish this task; the
differential and the integral approach. The differential approach
requires, in addition to distances, derivatives of distance with
respect to redshift as well as knowledge of the expansion history. The
integral approach is based on measuring the integral of the inverse
of the expansion history via measurements of cosmic time as derived, e.g.,
from galaxy ages.

In this paper, we attempt to constrain the large scale curvature of the
Universe using distances obtained from observations of Type Ia
supernovae together with inferred ages of passively evolving galaxies
and Hubble parameter estimates from the large scale clustering of
galaxies. Current data are consistent with zero spatial curvature,
although the uncertainty of $\ok$ is of order unity. Future data sets
with on the order of thousands of Type Ia supernovae distances and
galaxy ages will allow us to constrain the spatial curvature $\ok$
with an uncertainty of $\lesssim 0.1$ at the 95\% confidence level.} %abstract


\begin{document}
\maketitle
\flushbottom

%%%%%%%%%%%%%%%%%%%%%%%%%%%%%%%%%%%%%%%%%%%%%%%%%%%%%%%%%%%%%%%%%%%%%%%
%%%%%%%%%%%%%%%%%%%%%%%%%%%%%%%%%%%%%%%%%%%%%%%%%%%%%%%%%%%%%%%%%%%%%%%
\section{Introduction}
Current observational evidence is consistent with the universe being
made up of $\sim 5\%$ baryonic matter, $\sim 25\%$ non-baryonic
dark matter, and $\sim 70\%$ dark energy in the form of vacuum
energy, or equivalently, a cosmological constant
\cite{2009ApJS..180..330K,2009arXiv0908.4274K}. In this concordance
model, the total energy density is very close to the critical energy
density needed for the universe to have zero spatial curvature on
large scales. Considerable effort is put into probing the expansion
history of the universe in greater detail in order to pinpoint the
properties of the dominant dark energy component. Also, there are many
ongoing experiments aiming at observing the dark matter at particle
level. Less energy is being put into investigating the large scale
geometry of the universe.

Most inflationary models predict that the spatial curvature should be
on the order of $|\ok |\lesssim 10^{-5}$ (though there are models that
allow for larger curvature)
\cite{2006PhRvD..73b3503K,2009ApJS..180..330K}. 
As such, probing geometry at the largest scales can probe fundamental
physics. 

Traditionally, the angular size of the fluctuations in the cosmic
microwave background (CMB) has been regarded as the most effective way
to constrain spatial curvature. The common explanation is that for a
fixed physical size of the fluctuations, when compared to the case of
a flat geometry, positive curvature (i.e., $\ok<0$) will cause
the observed angle to be larger and negative curvature ($\ok>0$)
smaller. However, since the relation between the physical scale and
the observed angle is given by the angular diameter distance, not only
spatial geometry but also the expansion history will affect the
observed angle. Specifically, the reason why the observed angle is not
as sensitive to the matter density (as expected for a fixed physical
scale) is that the physical scale has a very similar dependence on the
matter density as the angular diameter distance and thus will be
factored out when calculating the observed angle. However, a non zero
$\ok$ affects both the spatial curvature and the expansion rate. Since
some of the sensitivity of the CMB observations is due to the effect
on the expansion rate, dark energy properties may also affect the
observed angular scale of the fluctuations.  Conversely, if the
spatial curvature of the Universe is non zero, it can mimic evolving
dark energy
\cite{cro03,pol05,2007JCAP...08..011C,hua07}. Because of this
degeneracy, measurements of the curvature parameter, $\ok$, usually
involve a parameterised description of the dark energy.  Recent
measurements of $\ok$, utilising combinations of different
cosmological probes and simple parameterisations of the dark energy
equation of state, indicate that the Universe is nearly flat
\cite{2009arXiv0910.0252B,gon05,ich06a,ich06b,wri07,zha07,rei09,vik09,2010arXiv1001.4538K}. 
However, some parameterisations allow large deviations from a flat
Universe \cite{ich07}. The relation between expansion and curvature
is also evident from how current bounds on the acceleration depends on
curvature \cite{2009JCAP...01..044M}.

The observed scales of CMB fluctuations are sensitive to the
combination $H_0^2\ok$, where $H_0$ is the Hubble constant. Assuming
that the dark energy is in the form of a cosmological constant and
incorporating Hubble Space Telescope (HST) priors on $H_0$, Wilkinson
Microwave Anisotropy Probe (WMAP) 5-year data
\cite{2009ApJS..180..330K} gives $-0.2851<\ok<0.0099$ at 95\%
confidence level (CL). If data from baryon acoustic oscillations (BAO)
and Type~Ia supernovae (SNe~Ia) are also included, the limit improves
to $-0.0181<\ok<0.0071$. With the latest WMAP 7-year data
\cite{2010arXiv1001.4538K}, these limits change only marginally to
$-0.0133<\ok<0.0084$.

Barenboim et al.~\cite{2009arXiv0910.0252B} investigate how
future BAO and CMB data can be used to resolve the degeneracy between
curvature and the dark energy equation of state, within a parameterised
model for $w(z)$.
% Knox
In \cite{2006PhRvD..73b3503K}, it is shown how $\ok$ can be measured
with high precision using future CMB data combined with distance
observations out to redshifts where the energy content of the universe
is dominated by matter. The distance measures can break the degeneracy
between $\ok$ and $\om$, and are less sensitive to the late time dark
energy properties.
% Mortonson
It was shown in \cite{mor09} how future observations of the growth of
structure in the universe could be used to constrain the expansion
history and thus, together with distance measurements from CMB and
SNe~Ia, can constrain $\ok$ to sub-percent accuracy.
% Bernstein
A method to probe the purely geometrical part of $\ok$ using weak
gravitational lensing and BAO was devised in
\cite{2006ApJ...637..598B}.
Note that, in contrast to the methods described in
\cite{2009arXiv0910.0252B,mor09,2006PhRvD..73b3503K}, this technique 
is completely independent of the expansion history of the universe and
thus to the dark matter and dark energy densities and properties.
This is also true for the methods described in this paper, utilising
data at lower redshifts. As such, they earn their merit as independent
probes of spatial curvature even though they will not be able to
compete with the (parameterised) constraints on $\ok$ from CMB
observations.

In section~\ref{sec:flrw} we outline the expansion history independent
methods employed in this paper. These methods rely upon observable
quantities introduced in section~\ref{sec:obs}. We use the available
data, described in section~\ref{sec:data}, to obtain constraints
presented in section~\ref{sec:results}. In section~\ref{sec:future},
future constraints on $\ok$ are investigated using simulated data
sets. The results are discussed and summarised in
section~\ref{sec:summary}.

%%%%%%%%%%%%%%%%%%%%%%%%%%%%%%%%%%%%%%%%%%%%%%%%%%%%%%%%%%%%%%%%%%%%%%%
%%%%%%%%%%%%%%%%%%%%%%%%%%%%%%%%%%%%%%%%%%%%%%%%%%%%%%%%%%%%%%%%%%%%%%%
\section{Disentangling cosmic curvature and expansion history}\label{sec:flrw}
In a Friedmann-Lema\^{i}tre-Robertson-Walker (FLRW) universe, the
expansion history at late times when the contribution from radiation
can be neglected, is given by
\begin{equation}
E(z) \equiv \frac{H(z)}{H_0}=\left[\om(1+z)^3 +\ok(1+z)^2+\Omega_{\rm DE}f(z)\right]^{1/2},
\label{eq:E}
\end{equation}
where $\om$, $\ok$ and $\Omega_{\rm DE}$ are the current (effective)
energy densities of matter, spatial curvature and dark energy,
respectively, in units of the current critical density. In terms of
the equation of state parameter, $w(z)$, the function governing the
evolution of the dark energy with time is
\begin{equation}
f(z)=\exp\left( 3\int_0^z\frac{1+w(x)}{1+x}dx\right). 
\end{equation} 
The parameterisation of the expansion history plays, however, no role
in our method.  Let us now define the quantity
\begin{equation}
I(z) \equiv \int_0^z \frac{dx}{E(x)},
\label{eq:I}
\end{equation}
which is proportional to the proper distance. We will make use of the
$H_0$-independent comoving distance
\begin{equation}
D(z) \equiv \frac{1}{\sqrt{-\ok}} \sin \left[ \sqrt{-\ok}I(z)  \right],
\label{eq:D}
\end{equation}
which is related to luminosity and angular diameter distances, $d_{\rm
L}$ and $d_{\rm A}$, through
\begin{equation}
D(z)=\frac{H_0 d_{\rm L}(z)}{(1+z)}=H_0d_{\rm A}(z)(1+z).
\end{equation}
We note that distances depend on the spatial curvature, $\ok$ in two
ways; it affects the expansion rate of the Universe as well as the
geometrical path of photons. Any effect on the expansion can be
mimicked by a dark energy component with equation of state 
$w=-1/3$, corresponding to $f(z)=(1+z)^2$. We therefore seek to isolate the purely geometrical effect of
curvature.

\begin{figure}[t]
\begin{center}
\includegraphics[angle=0,width=.49\textwidth]{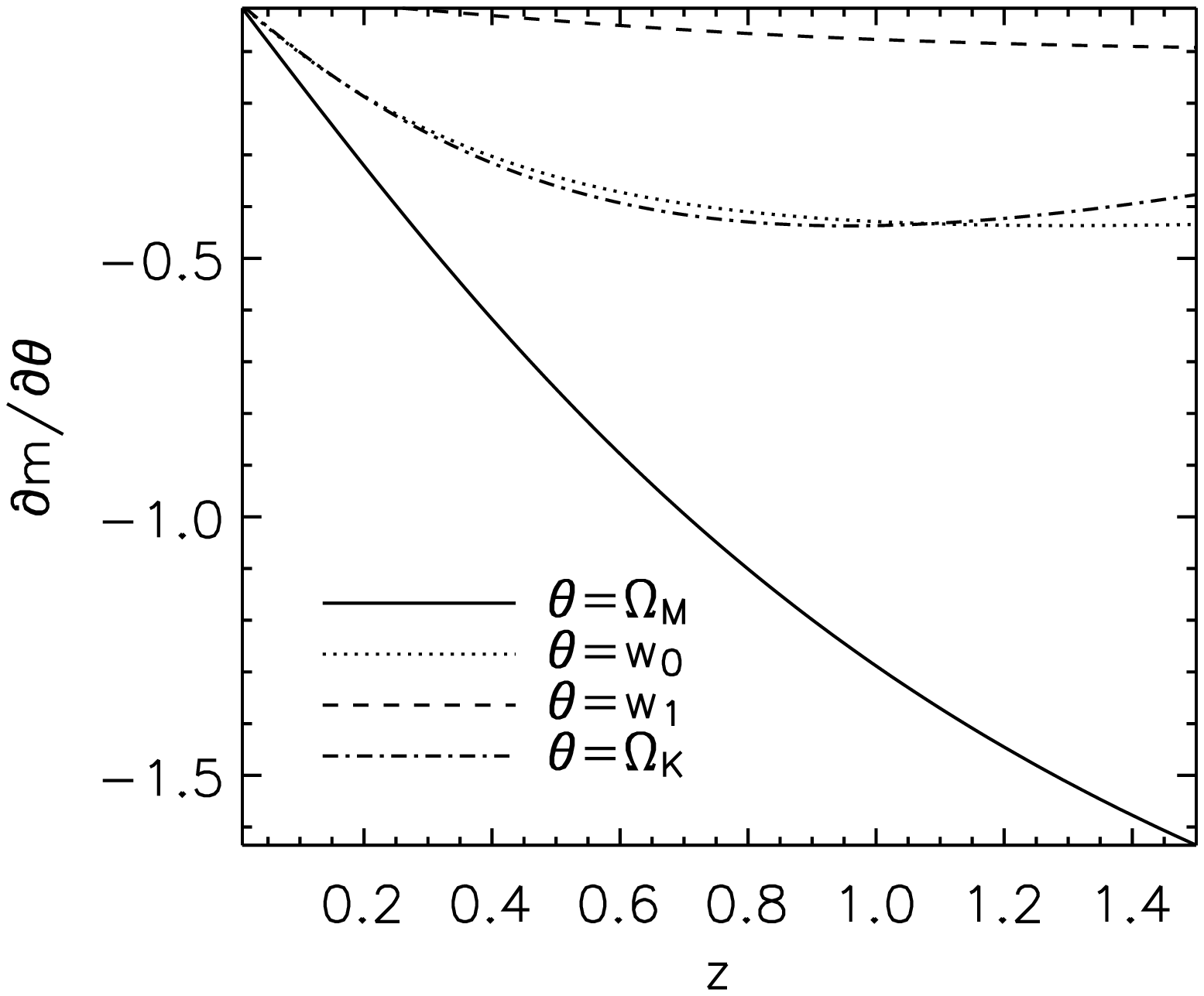}
\includegraphics[angle=0,width=.49\textwidth]{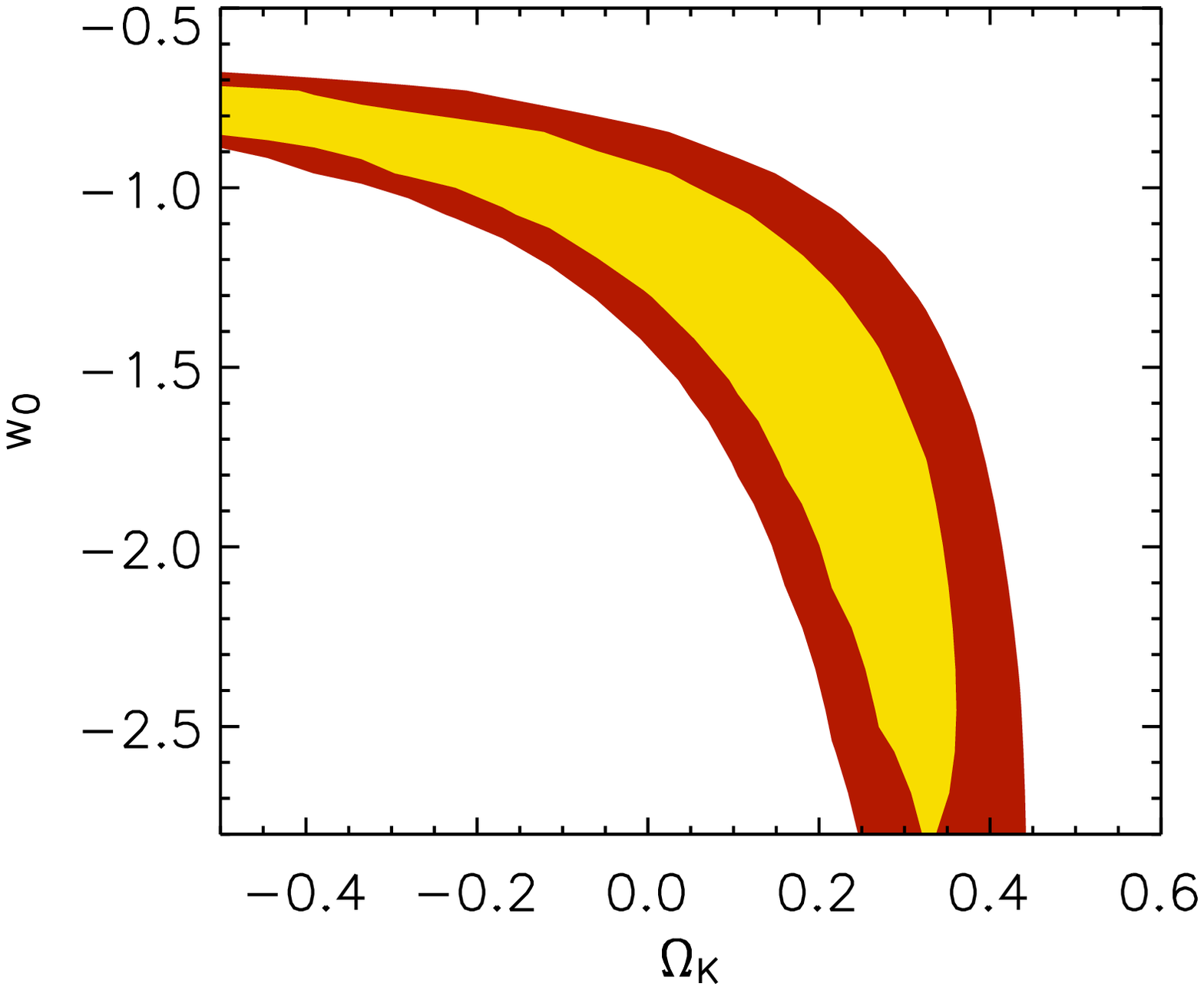}
\caption{\label{fig:dthetadm} {\em Left panel:} The sensitivity of the 
observed SN~Ia magnitudes to different cosmological parameters 
$\theta=\{\om,w_0,w_1,\ok\}$ as given by the derivatives
$\partial m/\partial w_0(z)$, $\partial m/\partial \ok(z)$ etc.  The cosmology is $\om=0.3, \ok=0,
w_0=-1, w_1=0$. {\em Right panel:} Contours at the 68.3 and 95\% CL in
the $[w_0,\ok]$-plane from SN~Ia observations. This is for the Union2
data set described in section~\ref{sec:sndata} assuming a fixed matter
density $\om=0.3$ and $w_1=0$.}
\end{center}
\end{figure}

The importance of not relying on any assumptions about the expansion
history can be appreciated by noting that we do not need very
elaborate dark energy models to mimic the effects from spatial
curvature on distances out to moderate redshifts. There is an almost
perfect degeneracy in distances already between a constant dark energy
equation of state parameter, $w_0$, and the curvature, $\ok$. In
figure~\ref{fig:dthetadm}, we show how the observed magnitude of a
source changes with redshift and cosmological parameters. We have used
a parameterisation of the dark energy equation of state given by
\begin{equation}
   w(z) = w_0+w_1\frac{z}{1+z}\, .
\end{equation}
Since the shapes of $\partial m/\partial w_0$ and $\partial m/\partial \ok$ are very similar,
confidence contours from the redshift-distance relation for SNe~Ia
(here we use the Union2 compilation of SN data described in
section~\ref{sec:sndata}) have a very degenerate structure when
fitting for $\ok$ and $w_0$, even for fixed values of $\om$ and
$w_1$. In a similar analysis using SDSS-II SN data
\cite{2009arXiv0908.4274K}, a non zero curvature is preferred
when using the MLCS2k2 light curve fitter and assuming that dark energy
is in the form of a cosmological constant
\cite{2009ApJ...703.1374S}.
%----------------------------------------------------------------------
%----------------------------------------------------------------------

Note that we assume that the spatial curvature is constant throughout
space. This would not be the case in, e.g., so called
Lema\^{i}tre-Tolman-Bondi models. However, it can be shown that in
such models, when combining SN~Ia data and CMB data, the observer need
to be uncomfortably close to the center, thus violating the Copernican
principle \cite{2009arXiv0909.4723B}. 
We also assume isotropy, i.e., that the universe looks the same in all
directions since neither the CMB nor SNe~Ia show any deviations from
this simple picture
\cite{2008JCAP...06..027B}.

From eq.~(\ref{eq:D}), it is clear that (luminosity and angular
diameter) distances are degenerate with respect to curvature and
expansion history. This degeneracy can be broken if the information
from distance measurements is supplemented by the expansion history,
$E(z)$, or the integral, $I(z)$. Two different approaches to measure
$\ok$ can therefore be pursued.

%%%%%%%%%%%%%%%%%%%%%%%%%%%%%%%%%%%%%%%%%%%%%%%%%%%%%%%%%%%%%%%%%%%%%%%
\subsection{Differential approach}
The \emph{differential} approach, originally outlined in
\cite{2007JCAP...08..011C}, is based on differentiating
eq.~(\ref{eq:D}). We can then write
\begin{equation}
\label{eq:ok}
   \ok=\frac{(ED')^2-1}{D^2},
\end{equation}
where primes denote derivatives with respect to redshift.  
This is the curvature contribution to the geometry only; the curvature
contribution to the expansion velocity is divided out by $E(z)$. The
differential approach requires not only distances, but also derivatives
of distances as well as independent estimates of $E(z)$.

Since $\ok$ is constant for any FLRW model,
eq.~(\ref{eq:ok}) measured at different redshifts would also
offer a test of the Copernican principle \cite{cla08}.  Similar
consistency relations for the concordance model, i.e.~a flat cosmological constant dominated universe, can be formed
by combining $E(z)$, $D(z)$ and its first and second derivatives
\cite{sha09}. 

%%%%%%%%%%%%%%%%%%%%%%%%%%%%%%%%%%%%%%%%%%%%%%%%%%%%%%%%%%%%%%%%%%%%%%%
\subsection{Integral approach}
Measurements of the integral $I(z)$ allows the \emph{integral}
approach to be followed. To gain insight into the relation between
cosmic curvature, distance, and $I(z)$ we can study a Taylor
expansion of eq.~(\ref{eq:D}) around a flat universe
\begin{equation}
  D\simeq I+\frac{1}{6}\ok I^3+\frac{1}{120}\ok^2 I^5+\mathcal{O}(\ok^3).
\label{eq:taylor}
\end{equation}
The distance $D$ is to first order equal to $I$ and $\ok$ determines
the magnitude of the higher order terms. Neglecting terms of order three and higher
in eq.~(\ref{eq:taylor}) leads to a second order equation with
the solution
\begin{equation}
\ok \simeq \frac{10}{I^2} \left\{ 
\sqrt{1+\frac{12}{10} \left[ \frac{D}{I}-1\right] }
-1
\right\}.
\label{eq:curv5}
\end{equation}
The accuracy of this approximation decreases for increasing
redshift. At $z=0.5$, 1.0, and 1.5 the error for $|\ok|<0.3$ is
smaller than 1, 2, and 6 parts in $10^5$, respectively.

%%%%%%%%%%%%%%%%%%%%%%%%%%%%%%%%%%%%%%%%%%%%%%%%%%%%%%%%%%%%%%%%%%%%%%%
%%%%%%%%%%%%%%%%%%%%%%%%%%%%%%%%%%%%%%%%%%%%%%%%%%%%%%%%%%%%%%%%%%%%%%%
\section{Observable quantities}\label{sec:obs}
In the previous section we described how the differential and integral
approach can be used to measure the large scale curvature of the
Universe in a model independent way. We now discuss how $D(z)$,
$D'(z)$, $E(z)$, and $I(z)$ can be obtained from observable
quantities.

%%%%%%%%%%%%%%%%%%%%%%%%%%%%%%%%%%%%%%%%%%%%%%%%%%%%%%%%%%%%%%%%%%%%%%
\subsection{Measuring $D(z)$ and $D'(z)$}\label{sec:sndist}
\begin{figure}[t]
\begin{center}
\includegraphics[angle=0,width=.49\textwidth]{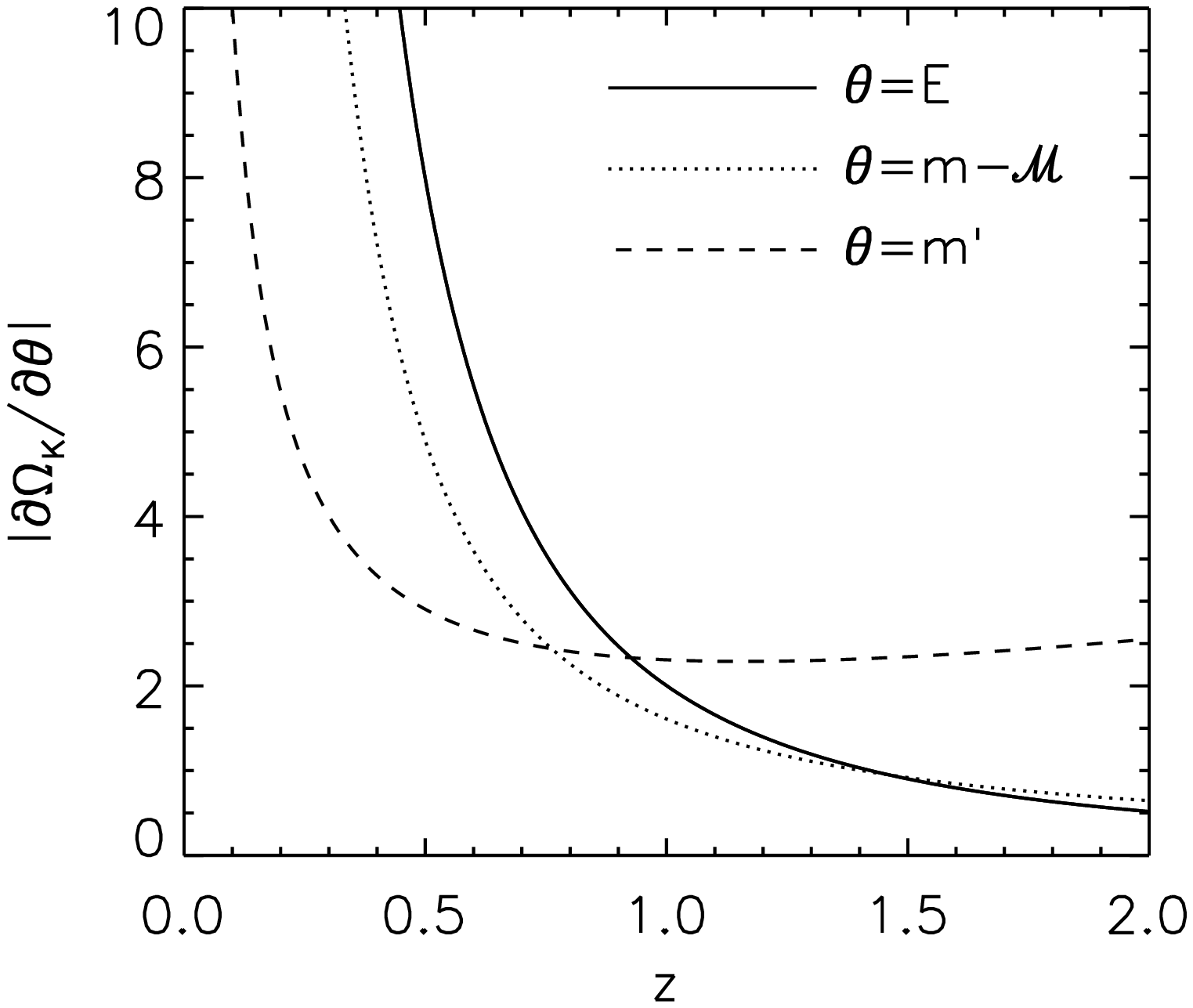}
\includegraphics[angle=0,width=.49\textwidth]{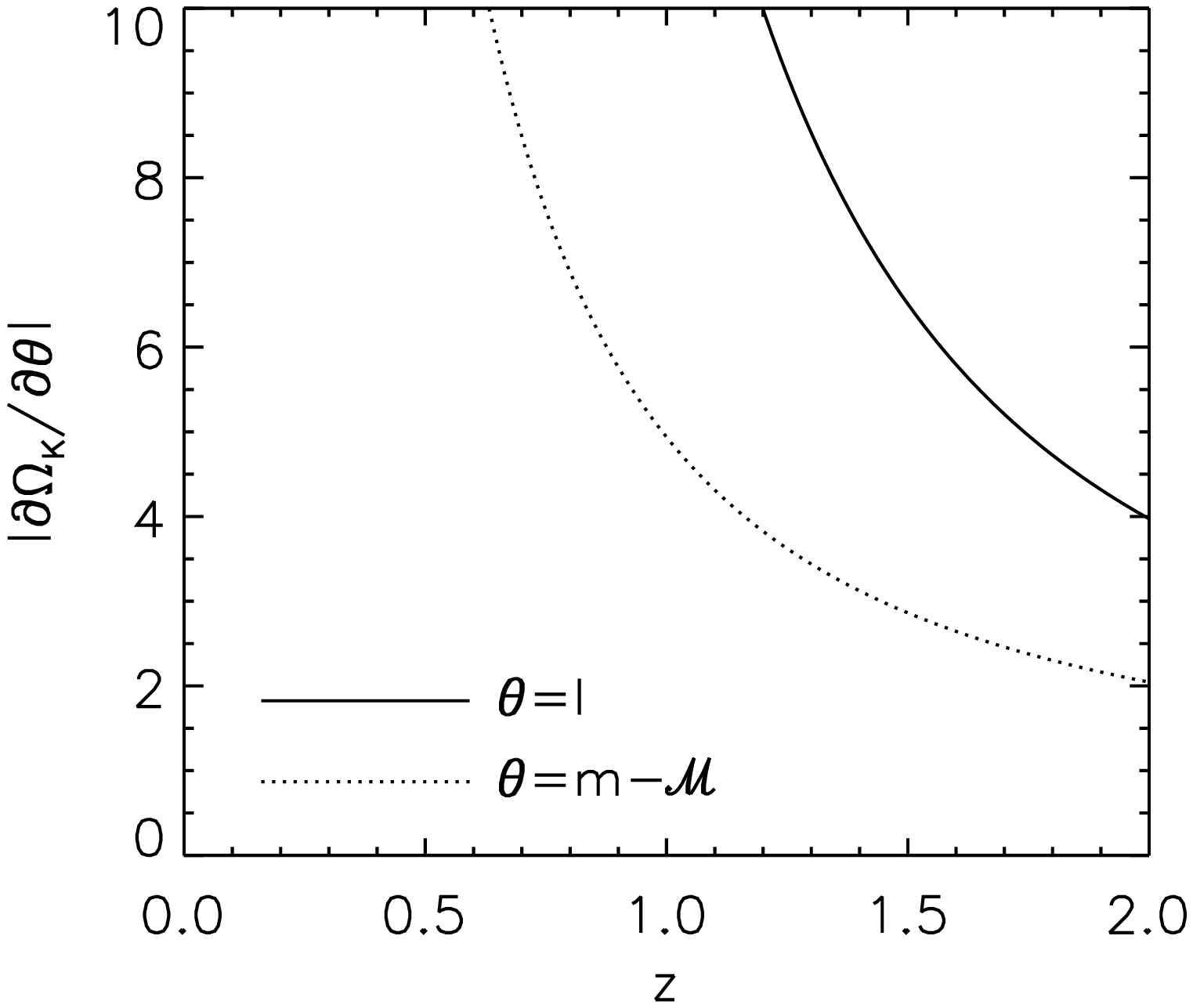}
\caption{\label{fig:derivs}  {\em Left panel:} 
Sensitivity of curvature computed using the differential approach to
different observables $\theta=\{E,m-\mathcal{M},m'\}$.  A high value
indicates that the observed quantity, e.g. $E(z)$ is not very
sensitive to the curvature and we expect the error of $\ok$ to be
large. {\em Right panel:} Sensitivity of curvature computed using the
integral approach to the observables
$\theta=\{I,m-\mathcal{M}\}$. The underlying cosmology was for both
plots assumed to be $\om=0.3$ and $\ola=0.7$.}
\end{center}
\end{figure}

The apparent magnitude, $m$, of a SN~Ia with absolute magnitude
$M$ is related to the distance $D$ through
\begin{equation}
   m=M + 5\log{\left(\frac{d_L}{\rm
   Mpc}\right)}+25\equiv\mathcal{M}+5\log{(1+z)}+5\log{D},
\end{equation}
where
\begin{equation}
   \mathcal{M}\equiv M+5\log\left({\frac{c/H_0}{1\, {\rm Mpc}}}\right)+25.
\end{equation}
This can be rewritten as
\begin{equation}
   m-\mathcal{M}= \frac{5}{\ln{10}}\left[\ln{(1+z)+\ln{D}}\right].
\end{equation}	
Consequently, the derivative of $m$ with respect to redshift is
\begin{equation}
   m'= \frac{5}{\ln{10}}\left[\frac{1}{(1+z)}+\frac{D'}{D}\right].
\end{equation}
These two relations can be inverted to yield an explicit formula for
the distance,
\begin{equation}
   D= \frac{1}{1+z}\exp{\left[\frac{\ln{10}}{5}(m-\mathcal{M})\right]},
      \label{eq:dsn}
\end{equation}
 and its derivative,
\begin{equation}
   D'= D\left[\frac{\ln{10}}{5}m'+\frac{1}{(1+z)}\right],
\end{equation}
in terms of observable quantities.
We can now rewrite eq.~(\ref{eq:ok}) as
\begin{equation}
   \ok=E^2\left[\frac{\ln{10}}{5}m'-\frac{1}{(1+z)}\right]^2
   -(1+z)^2\exp{\left[-0.4\ln{10}(m-\mathcal{M})\right]},
\end{equation}
i.e., $\ok=\ok (E,m-\mathcal{M},m')$. If we can measure
$E$, $m-\mathcal{M}$, and $m'$, we can thus constrain the spatial
curvature. The error on $\ok$ is given by
\begin{equation}
   \sigma_{\ok}=\sqrt{ \left(\frac{\partial\ok}{\partial E}\sigma_E\right)^2+
   \left(\frac{\partial \ok}{\partial(m-\mathcal{M})}\sigma_{m-\mathcal{M}}\right)^2+
   \left(\frac{\partial \ok}{\partial m'}\sigma_{m'}\right)^2}.
\end{equation}
In the left panel in figure~\ref{fig:derivs} we show how the sensitivity of $\ok$ to
the observables, $\theta=\{ E,m-\mathcal{M},m'\}$, varies with redshift, assuming a flat cosmological
constant universe with $\om=0.3$. We expect uncertainties in $E$ to
dominate at low redshifts and uncertainties in $m'$ to dominate
at $z\gtrsim 1$. Note that, since $\ok$ is constant, we can combine
data at different redshifts when constraining the global value of
$\ok$.

%----------------------------------------------------------------------
%----------------------------------------------------------------------

%%%%%%%%%%%%%%%%%%%%%%%%%%%%%%%%%%%%%%%%%%%%%%%%%%%%%%%%%%%%%%%%%%%%%%%
\subsection{Measuring $E(z)$}
One way to measure $E(z)$ is to use the evolution of lookback time with redshift.
The $H_0$-independent lookback time to an object at redshift $z$ is given by
\begin{equation}
\tau(z) \equiv H_0(t_0-t_z)=\int_0^z\frac{dx}{(1+x)E(x)}.
\label{eq:lookb}
\end{equation}
By differentiating eq.~(\ref{eq:lookb}) and rearranging terms, it is
clear that the expansion history can be obtained by differentiation of
the lookback time \cite{jim02},
\begin{equation}
E(z)=-\frac{1}{(1+z)}\frac{dz}{d\tau}.
\label{eq:dage}
\end{equation}
In appendix~\ref{app:lookb}, the relationship between lookback time
and galaxy age is discussed. According to eq.~(\ref{eq:iobs}), the
following equality holds
\begin{equation}
\frac{d \tau}{dz}=-H_0\frac{dt_{\rm age}}{dz},
\label{eq:tder}
\end{equation}
implying that the evolution of galaxy age, $t_{\rm age}(z)$, with
redshift can be used to measure the expansion history.

%%%%%%%%%%%%%%%%%%%%%%%%%%%%%%%%%%%%%%%%%%%%%%%%%%%%%%%%%%%%%%%%%%%%%%%
\subsection{Measuring $I(z)$}
Galaxy ages can also be used to measure the
integral~(\ref{eq:I}). Taking the derivative of eq.~(\ref{eq:lookb}),
rearranging terms and then integrating, we arrive at
\begin{equation}
I(z)=\int_0^z \frac{d \tau}{d z}(1+x)dx=\tau(z)(1+z)-\int_0^z \tau(x)dx,
\label{eq:iage}
\end{equation}
where integration by parts was used in the last step. 
Substituting galaxy ages for lookback times in eq.~(\ref{eq:iage})
comes at a price; an unknown constant $T$ enters in
eq.~(\ref{eq:iobs}), relating galaxy age and lookback time.

In order to compute the integral in eq.~(\ref{eq:iage}), we need to
measure $t_{\rm age}$ as a function of redshift. If measurements below
$z_{\rm min}$ are lacking, we have to take the integral
$\int_0^{z_{\rm min}}t_{\rm age}(x)dx$ into account. Fortunately
this constant can be added to $T$.  The integral $I$ is therefore
related to galaxy ages via
\begin{equation}
I(z)=H_0\left[\mathcal{I} 
-t_{\rm age}(z)(1+z)+\int_{z_{\rm min}}^z t_{\rm age}(x)dx
\right],
\label{eq:imeas}
\end{equation} 
where we have defined a new constant
\begin{equation}
\mathcal{I}=T+\int_0^{z_{\rm min}} t_{\rm age}(x)dx,
\end{equation}
which has to be determined from data. 

In terms of observable quantities, eq.~(\ref{eq:curv5}) can be rewritten as
\begin{equation}
\ok \simeq \frac{10}{I^2} \left\{ 
\sqrt{1+\frac{12}{10} \left[ \frac{ \exp \left[ \frac{\ln 10}{5}(m-\mathcal{M})\right]}{(1+z)I}-1\right] }
-1
\right\},
\label{eq:curv5b}
\end{equation}
i.e., $\ok=\ok(I, m-\mathcal{M})$.
The uncertainty in a measurement of the curvature using the integral approach is given by
\begin{equation}
\sigma_{\ok}=\sqrt{\left( \frac{\partial \ok}{\partial I} \sigma_I \right)^2+
\left( \frac{\partial \ok}{\partial (m-\mathcal{M})}\sigma_{m-\mathcal{M}} \right)^2}.
\end{equation}
The right panel in figure~\ref{fig:derivs} shows how the sensitivity of $\ok$ to the variables $\theta=\{ I,m-\mathcal{M} \}$ varies with redshift. The error decreases with redshift
and is dominated by the error in $I(z)$. 
Figure~\ref{fig:derivs} shows that for a given error in
$m-\mathcal{M}$, the uncertainty in $\ok$ is larger for the integral
than for the differential approach.

%%%%%%%%%%%%%%%%%%%%%%%%%%%%%%%%%%%%%%%%%%%%%%%%%%%%%%%%%%%%%%%%%%%%%%%
\subsection{Differentiation of data}\label{sec:diffdata}
The differential method relies on estimating derivatives of noisy
data, SN~Ia magnitudes to get $m'$ and, in the case we do not have
$H(z)$ data, the derivative of galaxy ages to obtain $E(z)$ (see
section~\ref{sec:age}).

Since, in a flat universe, $m'$ gives a measure of $H(z)$, SN~Ia data
has been used to show how $H(z)$ increases with redshift. A common way
to derive $m'$ for this purpose is to use the method described in
\cite{2005PhRvD..71j3513W} for extracting the expansion history in
uncorrelated redshift bins from SN~Ia data
\cite[e.g.,][]{2007ApJ...659...98R,2009JCAP...01..044M,2009ApJ...703.1374S}.
It turns out that this method is equivalent to the much simpler method
of fitting straight lines to the $m(z)$ data in redshift bins and use
the slope of the fitted lines as estimates of the derivative (with
corresponding errors). We have checked that fitting higher order
polynomials to the data in a given bin does not affect the estimated
slope appreciably\footnote{Note that this equivalent to saying that
when estimating the differential using a Savitsky-Golay filter of
degree $n$, the results is insensitive to $n$.}.

Having $n_{\rm SN}$ SNe~Ia with magnitude uncertainty $\sigma_m$ in a
redshift bin of width $\Delta z$ gives an uncertainty in $m'$ of
\begin{eqnarray}
  \sigma_{m'}&\simeq &\frac{\sigma_m}{0.2886\Delta z\sqrt{n_{\rm SN}-2}}\\
  &\simeq &0.031\left(\frac{n_{\rm SN}}{500}\right)^{-1/2}
    \left(\frac{\Delta z}{0.5}\right)^{-1}\left(\frac{\sigma_m}{0.1}\right).
\end{eqnarray}
This can be compared to the error in $m-\mathcal{M}$ given by
\begin{equation}
  \sigma_{m-\mathcal{M}}=\sqrt{\sigma_{m}^2+\sigma_{\mathcal{M}}^2},
\end{equation}
where
\begin{equation}
  \sigma_{m}\simeq 0.0045\left(\frac{n_{\rm SN}}{500}\right)^{-1/2}
    \left(\frac{\sigma_m}{0.1}\right).
\end{equation}
and $\sigma_{\mathcal{M}}$ is given by an equivalent formula where
$n_{\rm SN}$ refers to the number of low redshift SNe Ia.

Estimating $H(z)$ from $n_{\rm age}$ galaxy age estimates with
fractional uncertainty $f_{\rm age}$ in a redshift bin of width
$\Delta z$, we obtain
\begin{equation}
  \frac{\sigma_{H}}{H}\simeq
  0.031(1+z)\left(\frac{n_{\rm age}}{500}\right)^{-1/2} \left(\frac{\Delta
  z}{0.5}\right)^{-1}\left(\frac{f_{\rm age}}{0.1}\right)\, .
\end{equation}
Besides the error in $H(z)$, we must also consider the error in the
Hubble constant when estimating the error in $E(z)$,
\begin{equation}
  \sigma_{E}=E\sqrt{\left(\frac{\sigma_{H}}{H}\right)^2+\left(\frac{\sigma_{H_0}}{H_0}\right)^2}.
  \label{eq:eerr}
\end{equation}
When combining results from different redshift bins, one must take
into account the correlations between the bins since any error in
$\mathcal{M}$ and $H_0$ will cause a systematic shift in the derived
value of $\ok$.

%%%%%%%%%%%%%%%%%%%%%%%%%%%%%%%%%%%%%%%%%%%%%%%%%%%%%%%%%%%%%%%%%%%%%%%
\subsection{Integration of data}
Let us assume that we have measured a set of galaxy ages 
and their redshifts.
The following formula can then be used to compute the integral of $t_{\rm age}(z)$ numerically
\begin{equation}
F(z_k) \equiv \int_{z_{\rm min}}^{z_k} t_{\rm age}(x)dx \simeq \frac{1}{2}\sum_{j=1}^{k} 
(t_{{\rm age},j}+t_{{\rm age},j-1})(z_j-z_{j-1}),
\label{eq:int}
\end{equation}
where $z_k>z_j$ if $k>j$ and we assume $z_0=z_{\rm min}$.  Values of
$F(z_k)$ at different redshifts computed using eq.~(\ref{eq:int}) are
correlated. In appendix~\ref{app:err} we compute the covariance
matrix, $U_{jk}$, for a model where the data points are equidistant in
redshift, i.e.~the number density is constant. We also assume that the
errors are the same for all measurements. For the covariance matrix
given by eq.~(\ref{eq:cov}), we find that $U_{jk}
\lesssim U_{kk}$, i.e.,~the off diagonal elements are comparable in
magnitude to the variance $\sigma_F^2 \equiv U_{kk}$. However, when
computing $I(z)$ using eq.~(\ref{eq:imeas}) other sources of error
enter
\begin{equation}
\sigma_I=H_0\sqrt{\left( \frac{I}{H_0} \right)^2 \left( \frac{\sigma_{H_0}}{H_0} \right)^2
+\sigma^2_{\mathcal{I}}+(1+z)^2\sigma^2_{t_{\rm age}}+\sigma_F^2},
\end{equation}
where $\sigma_{\mathcal{I}}$ is the error in $\mathcal{I}$. The two last terms 
depends on $t_{\rm age}$ and we should therefore compare the covariance matrix
 $U_{jk}$
with the variance $(1+z)^2\sigma_{t_{\rm age}}^2$.  For our model the
covariance is inversely proportional to the number density and can
therefore be beaten down by statistics. If we approximate the
covariance with eq.~(\ref{eq:var}), $U_{jk}$ will be smaller than
$(1+z)^2\sigma_{t_{\rm age}}^2$ as long as the constant number density
$\alpha\equiv dn/dz$ is
\begin{equation}
\alpha > \frac{1}{4} \ge \frac{z-z_{\rm min}}{(1+z)^2}.
\end{equation}
Let us as an example take $\alpha=25$ corresponding to $n_{\rm age}
\simeq 40$ galaxy ages evenly distributed in the redshift interval $
0 \lesssim z \lesssim 1.5$. In that case $U_{jk}$ will be smaller than
$(1+z)^2\sigma_{t_{\rm age}}^2$ by a factor 100. In the following we
will therefore assume that the error in the integral approximation can
be neglected.

%%%%%%%%%%%%%%%%%%%%%%%%%%%%%%%%%%%%%%%%%%%%%%%%%%%%%%%%%%%%%%%%%%%%%%%
%%%%%%%%%%%%%%%%%%%%%%%%%%%%%%%%%%%%%%%%%%%%%%%%%%%%%%%%%%%%%%%%%%%%%%%
\section{Data}\label{sec:data}

%%%%%%%%%%%%%%%%%%%%%%%%%%%%%%%%%%%%%%%%%%%%%%%%%%%%%%%%%%%%%%%%%%%%%%
\subsection{Type Ia supernovae}\label{sec:sndata}
We use the Union2 \cite{2010ApJ...716..712A} compilation of SNe~Ia,
which is an updated version of the Union data set
\cite{2008arXiv0804.4142K}. This data set contains SNe~Ia from, e.g.,
the Supernova Legacy Survey, ESSENCE survey and HST
observations. After selection cuts, the data set amounts to 557
SNe~Ia, spanning a redshift range of $0\lesssim z \lesssim 1.4$,
analysed in a homogeneous fashion using the spectral-template-based fit
method SALT2.

%%%%%%%%%%%%%%%%%%%%%%%%%%%%%%%%%%%%%%%%%%%%%%%%%%%%%%%%%%%%%%%%%%%%%%
\subsection{Galaxy ages}\label{sec:age}
In passively evolving galaxies the stellar population was formed at
high redshift and has since then been evolving without any further
episodes of star formation. Massive galaxies located in high density
regions of clusters have old stellar populations and are therefore
believed to be passively evolving
\cite{dun96,spi97,cow99,hea04,tho05,tre05,pan07}. 
The ages of the passively evolving galaxies can be inferred from their
spectra using synthetic stellar population models
\cite{sto95,dun96,spi97,sto01}. Spectroscopic dating of galaxies have 
been used for cosmological purposes in, e.g.,
\cite{alc01,jim02,jim03,2005PhRvD..71l3001S,2010JCAP...02..008S}.

We use values of $H(z)$ derived from measured galaxy ages in 
\cite{2010JCAP...02..008S} for the differential approach 
(though look at the caveats in section~\ref{sec:summary}). For the
integral approach we use measurements of galaxy ages from
\cite{2005PhRvD..71l3001S}.

%%%%%%%%%%%%%%%%%%%%%%%%%%%%%%%%%%%%%%%%%%%%%%%%%%%%%%%%%%%%%%%%%%%%%%%
\subsection{Other probes of the Hubble parameter}
The Hubble parameter, $H(z)$, can in principle be measured with other probes
than galaxy ages, e.g., the time drift of redshifts
\cite{san62} and the dipole of the luminosity distance
\cite{bon06}. 

In \cite{2009MNRAS.399.1663G}, a peak along the radial direction of
the 2-point correlation function of LRG galaxies, consistent with the
expected BAO signal was found. Using the BAO peak position as a
standard ruler in the radial direction, enables a direct measurement
of $H(z)$ either at $z=0.24$ and $z=0.43$ or at $z=0.34$.  See,
however, also \cite{2009arXiv0901.1219M,2010arXiv1011.2729C} 
for discussions of the validity of the
results.

For the Hubble constant, throughout this paper, we use $H_0=74.2 \pm
3.6$ km s$^{-1}$ Mpc$^{-1}$ as derived from the HST distance ladder
observations of Cepheid variables, SNe Ia, and masers \cite{rie09}.

%%%%%%%%%%%%%%%%%%%%%%%%%%%%%%%%%%%%%%%%%%%%%%%%%%%%%%%%%%%%%%%%%%%%%%%
%%%%%%%%%%%%%%%%%%%%%%%%%%%%%%%%%%%%%%%%%%%%%%%%%%%%%%%%%%%%%%%%%%%%%%%
\section{Results}\label{sec:results}

%%%%%%%%%%%%%%%%%%%%%%%%%%%%%%%%%%%%%%%%%%%%%%%%%%%%%%%%%%%%%%%%%%%%%%%
\subsection{Differential approach}\label{sec:resdiffmeth}
\begin{figure}[t]
\begin{center}
\includegraphics[angle=0,width=.95\textwidth]{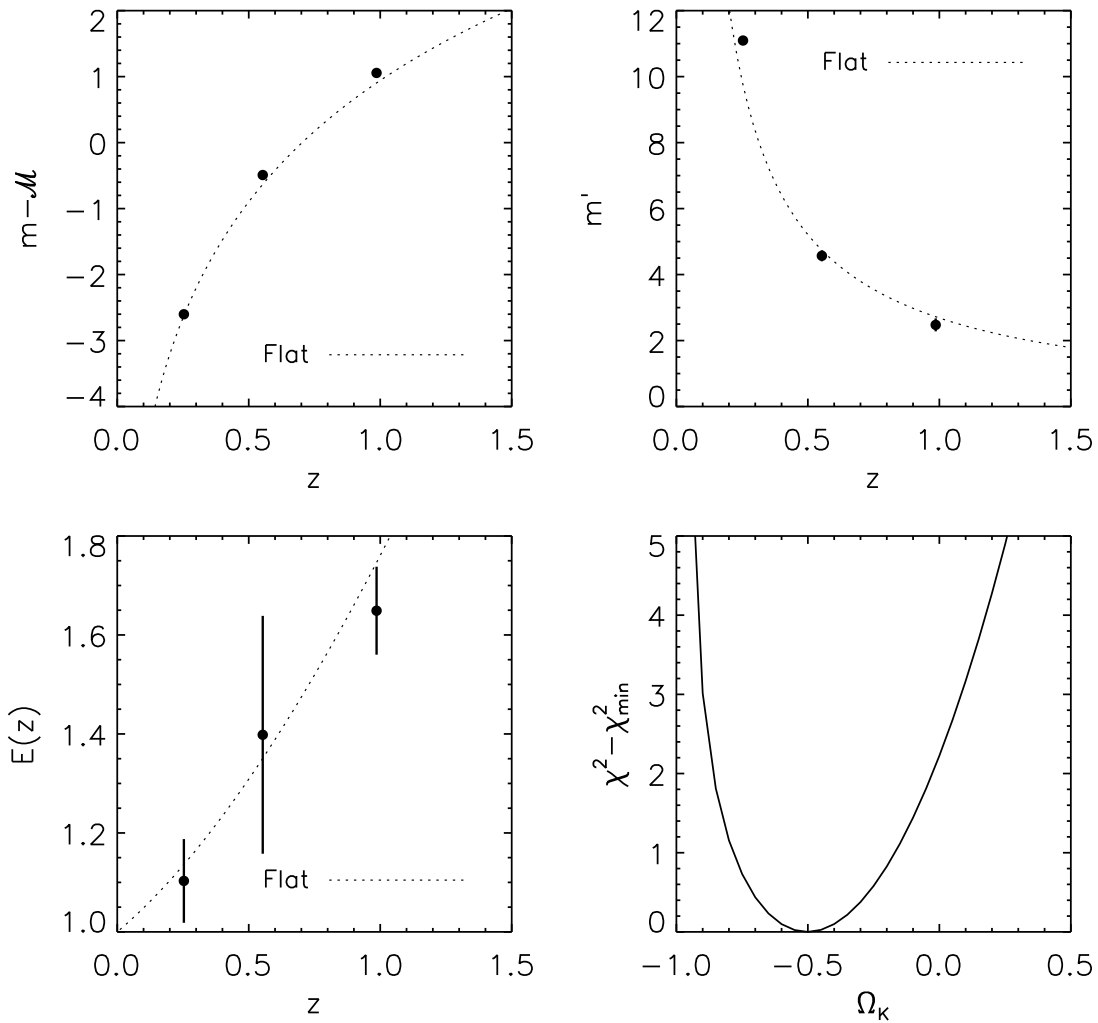}
\caption{\label{fig:stern} {\em Upper left:} $m-\mathcal{M}$ for the Union2 data set.  
Note that the error bars are smaller than the size of the plotting
symbols. {\em Upper right:} $m'$ for the Union2 data set. {\em Lower
left:} Values of $E(z)$ derived from galaxy ages in
\cite{2010JCAP...02..008S} and HST Hubble constant data \cite{rie09}.
The dotted lines correspond to the theoretical predictions
for a flat cosmological constant universe with $\om=0.3$. 
{\em
Lower right:} $\chi^2-\chi^2_{\rm min}$ as a function of $\ok$ giving 
$\ok=-0.50^{+0.66}_{-0.41}$ (95\% CL).}
\end{center}
\end{figure}

Combining the Union2 SN~Ia data set with values of $H(z)$ derived from
galaxy ages in \cite{2010JCAP...02..008S} (again, note the caveats in
section~\ref{sec:summary}) and HST Hubble constant constraints, gives
the result presented in figure~\ref{fig:stern}. 
We have collected the data in three redshift
bins: $0.1 \le z < 0.4$, $0.4 \le z < 0.8$, and $z \ge 0.8$. Data
below $z=0.1$ are being used to constrain $\mathcal{M}$.

The upper panels show $m-\mathcal{M}$ and $m'$ as derived from the
Union2 data set. The lower left panel shows $E(z)$ as derived from
galaxy ages \cite{2010JCAP...02..008S} with the HST value for $H_0$
\cite{rie09}.
The dotted lines correspond to the theoretical predictions
for a flat cosmological constant universe with $\om=0.3$. 
In order to put the data points at the same redshift, we have shifted
each data point according to the expected shift given a specific
cosmology where the curvature is a free parameter and the matter
density is held fixed at $\om=0.3$. Note however that the results are not
sensitive to this shifting, nor to the value of $\om$ since the shift
is performed over a relatively narrow redshift range. In each redshift
bin, we perform a $\chi^2$-minimisation to find the best-fitting value
of $\ok$. The combined constraint on $\ok$ from all bins (taking the
correlation between the bins properly into account, see
section~\ref{sec:diffdata}) is shown in the lower right panel, giving
$\ok=-0.50^{+0.66}_{-0.41}$ (95\% CL).
%----------------------------------------------------------------------
%----------------------------------------------------------------------

\begin{figure}[t]
\begin{center}
\includegraphics[angle=0,width=.95\textwidth]{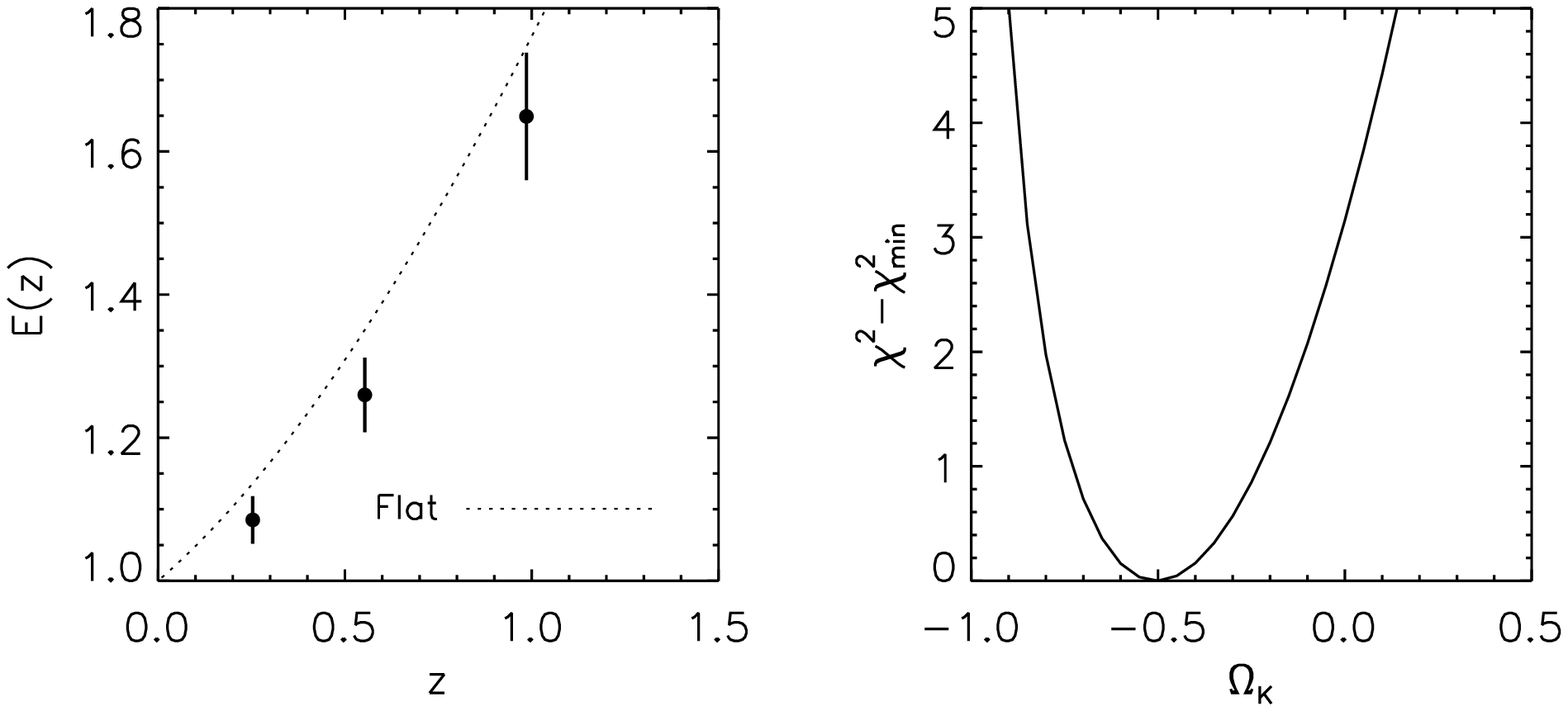}
\caption{\label{fig:fit2} {\em Left:} Values of $E(z)$ derived from galaxy ages in
\cite{2010JCAP...02..008S}, radial BAO constraints from \cite{2009MNRAS.399.1663G}, 
and HST Hubble constant data \cite{rie09}.  The dotted line
corresponds to the theoretical prediction for a flat cosmological
constant universe with $\om=0.3$. {\em Right:} $\chi^2-\chi^2_{\rm
min}$ as a function of $\ok$ giving $\ok=-0.50^{+0.54}_{-0.36}$ (95\%
CL).}
\end{center}
\end{figure}

Adding $H(z)$ derived from BAO data to the galaxy age data leads to
slightly smaller uncertainties on the expansion history $E(z)$ and
thus also the curvature: 
$\ok=-0.50^{+0.54}_{-0.36}$ (95\% CL), see figure~\ref{fig:fit2}.
%----------------------------------------------------------------------

%%%%%%%%%%%%%%%%%%%%%%%%%%%%%%%%%%%%%%%%%%%%%%%%%%%%%%%%%%%%%%%%%%%%%%%
\subsection{Integral approach}\label{sec:resintmeth}
\begin{figure}[t]
\begin{center}
\includegraphics[angle=0,width=.95\textwidth]{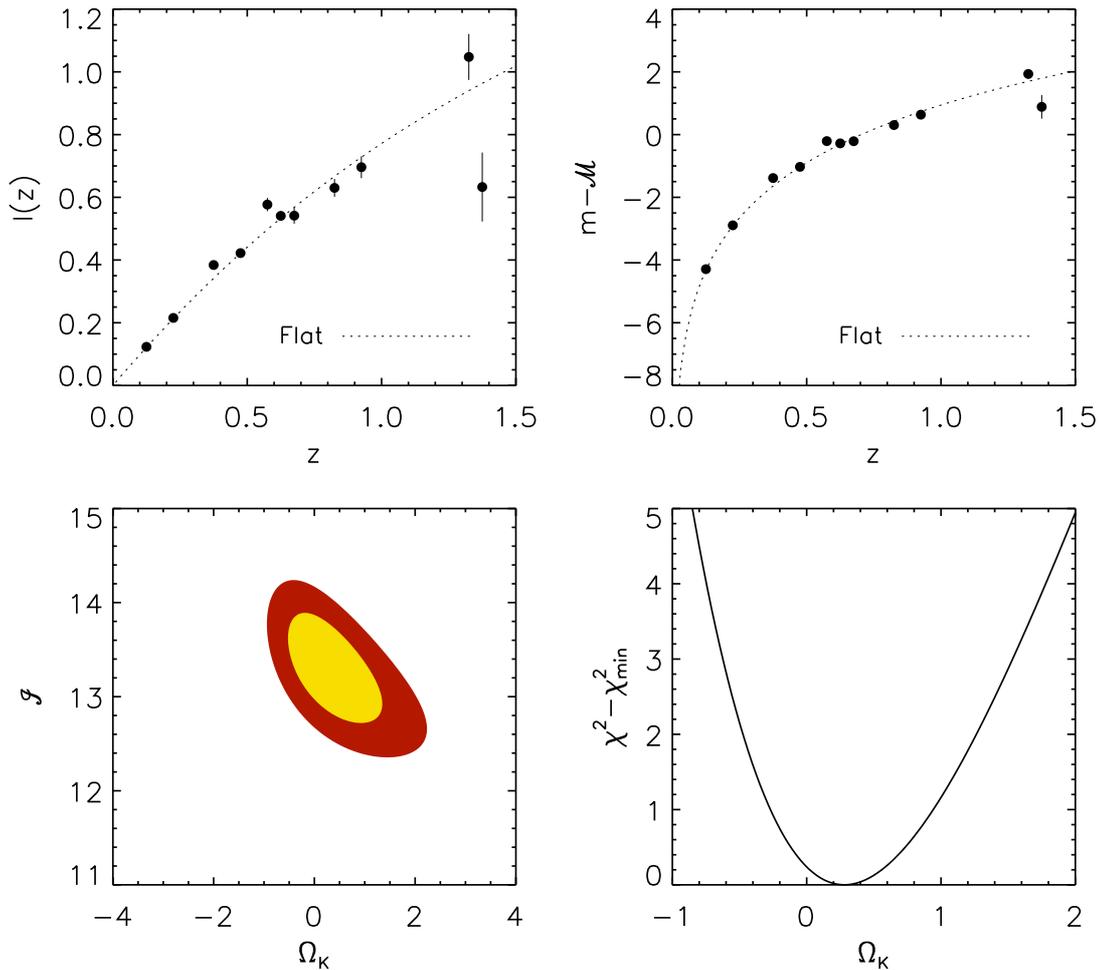}
\caption{\label{fig:fit} {\em Upper left:} $I(z)$ derived in redshift bins 
using galaxy ages from
\cite{2005PhRvD..71l3001S}. The best-fitting value of $\mathcal{I}$
and $H_0=74.2$ km s$^{-1}$ Mpc$^{-1}$ were used to compute $I(z)$
using eq.~(\ref{eq:imeas}).  {\em Upper right:} $m-\mathcal{M}$ for
the same redshift bins computed using the Union2 data
\cite{2010ApJ...716..712A} and a value of $\mathcal{M}$ fitted to low
redshift data.  {\em Lower left:} Confidence contours in the $[\ok,
\mathcal{I}]$-plane at the 68.3 and 95\% CL after marginalisation over
$H_0$ assuming a Gaussian prior. {\em Lower right:}
$\chi^2-\chi^2_{\rm min}$  as a function of $\ok$ giving 
$\ok=0.29^{+1.65}_{-0.94}$ (95\% CL).}
\end{center}
\end{figure}

We use galaxy ages from \cite{2005PhRvD..71l3001S} to estimate the
integral $I(z)$ as a function of redshift. The relative error in
$t_{\rm age}$ is assumed to be 10\% (L. Verde, private
communication). 
Computing $I(z)$ via eq.~(\ref{eq:imeas}) requires knowledge of $H_0$
and $\mathcal{I}$. The value of $\mathcal{I}$ depends on the redshift
range of the galaxy ages and when the star formation in the passively
evolving galaxies halted, it must therefore be inferred from the data
themselves.  Since $\mathcal{I}$ has to be determined from data, we fit
both $\ok$ and $\mathcal{I}$ to the data using
eq.~(\ref{eq:taylor}) instead of computing $\ok$ with
eq.~(\ref{eq:curv5b}). The upper left panel in figure~\ref{fig:fit}
shows $I(z)$ computed for the best-fitting value of $\mathcal{I}$
assuming $H_0=74.2$ km s$^{-1}$ Mpc$^{-1}$.  

Since SNe~Ia and galaxy ages are not measured at the same redshifts,
the data are collected in redshift bins of size $\Delta z=0.05$. For
the SNe~Ia, we require at least 3 SNe per bin when computing the
average magnitude, $\langle m \rangle$. In bins with more than one
value of $t_{\rm age}$ we use the average $\langle (1+z)t_{{\rm
age}}-F(z) \rangle$. The binning procedure results in $n_{\rm bin}=11$
bins ranging from $z_{\rm bin}=0.13$ to $1.38$. The upper right panel
in figure~\ref{fig:fit} shows the binned values of $m-\mathcal{M}$.

Results are also sensitive to the Hubble constant. We therefore
include $H_0$ in the fit and then marginalise over it assuming a
Gaussian prior to obtain constraints on $\ok$ and $\mathcal{I}$.
Using the binned data, we fit $\ok$, $\mathcal{I}$, and $H_0$ to the
data using the following $\chi^2$-statistic
\begin{equation}
\chi^2=\sum_{j=1}^{n_{\rm bin}}\frac{\left[D_j-
\left(I_j+\frac{\ok}{6}I_j^3+\frac{\ok^2}{120}I_j^5\right) \right]^2}
{\sigma_{D_j}^2+\left(1+\frac{\ok}{2}I_j^2+\frac{\ok^2}{24}I_j^4 \right)^2\sigma_{I_j}^2}, 
\label{eq:chi2}
\end{equation}
where $D_j=(1+z_{{\rm bin},j})^{-1}\exp \left[ \frac{\ln
10}{5}(\langle m\rangle_j-\mathcal{M})\right]$ and $I_j=H_0 \left[
\mathcal{I}-\langle (1+z)t_{\rm age}+F(z)\rangle_j \right]$. 
The errors $\sigma_{D_j}$ and $\sigma_{I_j}$ are computed using error
propagation from the measurements uncertainties in SN~Ia magnitudes
and galaxy ages, respectively.  

The lower left panel in figure~\ref{fig:fit} shows the confidence
levels in the $[\ok$,$\mathcal{I}]$-plane at 68.3 and 95\% CL. These
constraints were obtained after marginalisation over the Hubble
constant using the results of Riess et al.~\cite{rie09} as a Gaussian prior.  
Again we use a fixed value of
$\mathcal{M}$ determined from low redshift SNe~Ia.

The lower right panel in figure~\ref{fig:fit} shows
$\chi^2-\chi^2_{\rm min}$ as a function of $\ok$ after marginalisation over
$\mathcal{I}$. Our constraints on the curvature is
$\ok=0.29^{+1.65}_{-0.94}$ (95\% CL).

%----------------------------------------------------------------------
%----------------------------------------------------------------------

%%%%%%%%%%%%%%%%%%%%%%%%%%%%%%%%%%%%%%%%%%%%%%%%%%%%%%%%%%%%%%%%%%%%%%%
%%%%%%%%%%%%%%%%%%%%%%%%%%%%%%%%%%%%%%%%%%%%%%%%%%%%%%%%%%%%%%%%%%%%%%%
\section{Future}\label{sec:future}
Current data clearly give quite weak constraints on $\ok$.
In this section, we investigate the possibility to constrain the
spatial curvature using future distance and age, or $H(z)$, data. As
our default future data set, we assume that we will have 2000 well
observed SNe~Ia in the range $ 0.002 \le z \le 1.5$ with
$\sigma_m=0.1$ mag. For passive galaxy ages, we assume 2000
measurements in the range $ 0.1 \le z \le 1.5$ with a precision of 0.1
Gyr. Since these numbers are subject to large uncertainties, we also
discuss how our results scale with the total number and redshift
distribution of the data. If not otherwise stated, we assume that
$H_0$ has an uncertainty of one percent. Our fiducial cosmology is a flat 
universe with $\om=0.3$ and $\Omega_{\Lambda}=0.7$.

%%%%%%%%%%%%%%%%%%%%%%%%%%%%%%%%%%%%%%%%%%%%%%%%%%%%%%%%%%%%%%%%%%%%%%%
\subsection{Differential approach}
\begin{figure}[t]
\begin{center}
\includegraphics[angle=0,width=.95\textwidth]{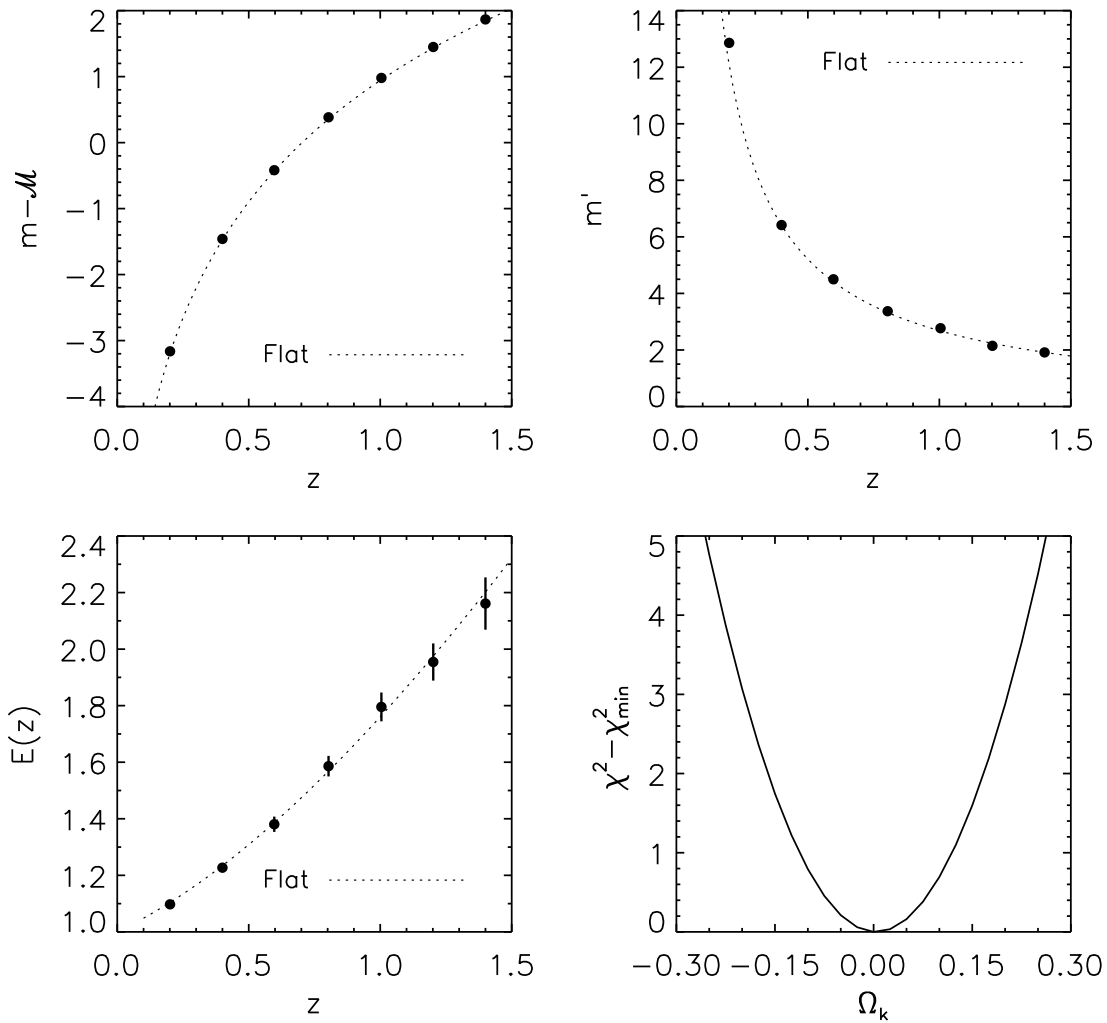}
\caption{\label{fig:smt2} Future constraints on $\ok$ with 
2000 well observed SNe~Ia in the range $ 0.002 \le z \le 1.5$ with
$\sigma_m=0.1$ mag and 2000 galaxy age measurements in the range $ 0.1
\le z \le 1.5$ with a precision of 0.1 Gyr. 
{\em Upper left:} $m-\mathcal{M}$ as derived from SN Ia
distances. {\em Upper right:} $m'$ as derived from SN Ia
distances. {\em Lower left:} $E(z)$ derived from galaxy ages. The
dotted lines are for a flat cosmology with $\om =0.3$ and a
cosmological constant. {\em Lower right:} The resulting
$\chi^2-\chi^2_{\rm min}$ giving $\ok=0.003 \pm 0.23$ (95\% CL).}
\end{center}
\end{figure}

For our default future data set, the differential method gives the
results depicted in figure~\ref{fig:smt2}, where we have $\ok=0.003 \pm
0.23$ (95\% CL). In addition, we have assumed that we have on the
order of 1000 SNe~Ia at low redshifts in order to calibrate
$\mathcal{M}$, although the error contribution would be negligible
also for a much smaller number of low redshift SNe.

\begin{figure}[t]
\begin{center}
\includegraphics[angle=0,width=.95\textwidth]{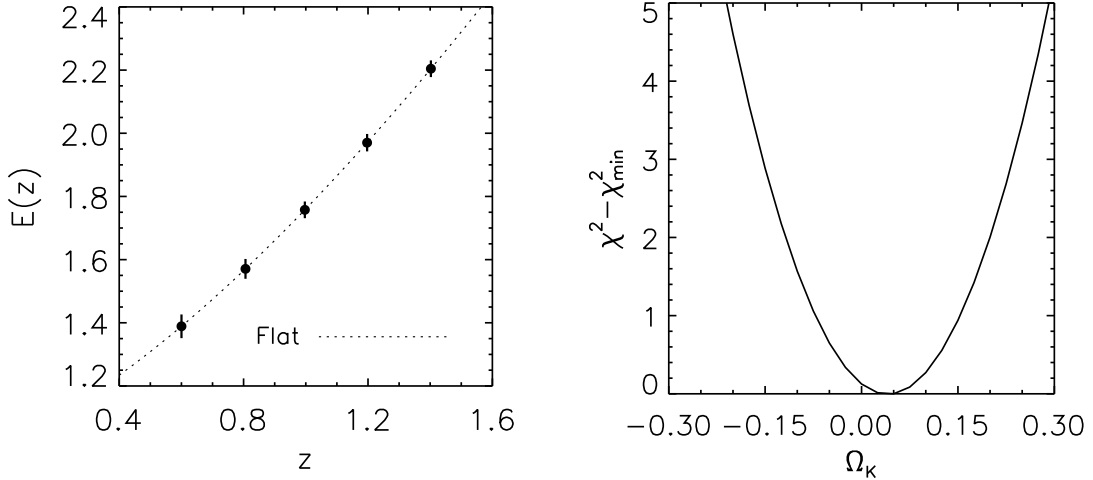}
\caption{\label{fig:sm2} Future constraints on $\ok$ with 
2000 well observed SNe~Ia in the range $ 0.002 \le z \le 1.5$ with
$\sigma_m=0.1$ mag and constraints on $H(z)$ at the few percent level
in redshift bins of size $\Delta z=0.2$ in the redshift interval
$0.5<z<1.5$ from future galaxy surveys
\cite{2005ApJ...631....1G}. {\em Left:} $E(z)$ as derived from the
radial BAO signal in future galaxy surveys and Hubble constant
constraints. The dotted line is for a flat cosmology with $\om =0.3$
and a cosmological constant. {\em Lower right:} The resulting
$\chi^2-\chi^2_{\rm min}$ giving $\ok=0.04 \pm 0.22$ (95\% CL).}
\end{center}
\end{figure}

This error on $\ok$ is comparable to what one can get with future
percent level bounds on the expansion rate, e.g. from radial BAO
measurements, instead of the galaxy ages.  In
\cite{2005ApJ...631....1G}, it is shown that a future large
high-redshift spectroscopic galaxy survey covering 10\,000 square
degrees could give a measure of $H(z)$ at the few percent level in
redshift bins of size $\Delta z=0.2$ in the redshift interval
$0.5<z<1.5$ by measuring the radial BAO signal. Combining these data
with our default SN Ia data set, we typically obtain results as
depicted in figure~\ref{fig:sm2}, where we have $\ok=0.04 \pm 0.22$
(95\% CL).

We note that our error scales as the inverse of the square root of the
number of SNe and/or galaxy ages, depending on which dominates the
error budget. For our default data set, the error contribution from
$E(z)$ and $m'$ are comparable at low redshifts whereas at high
redshifts, errors from $m'$ dominate. Generally, the errors from
$m-\mathcal{M}$ are negligible.

%----------------------------------------------------------------------
%----------------------------------------------------------------------

%----------------------------------------------------------------------
%----------------------------------------------------------------------

%%%%%%%%%%%%%%%%%%%%%%%%%%%%%%%%%%%%%%%%%%%%%%%%%%%%%%%%%%%%%%%%%%%%%%%
\subsection{Integral approach}
\begin{figure}[t]
\begin{center}
\includegraphics[angle=0.,width=.95\textwidth]{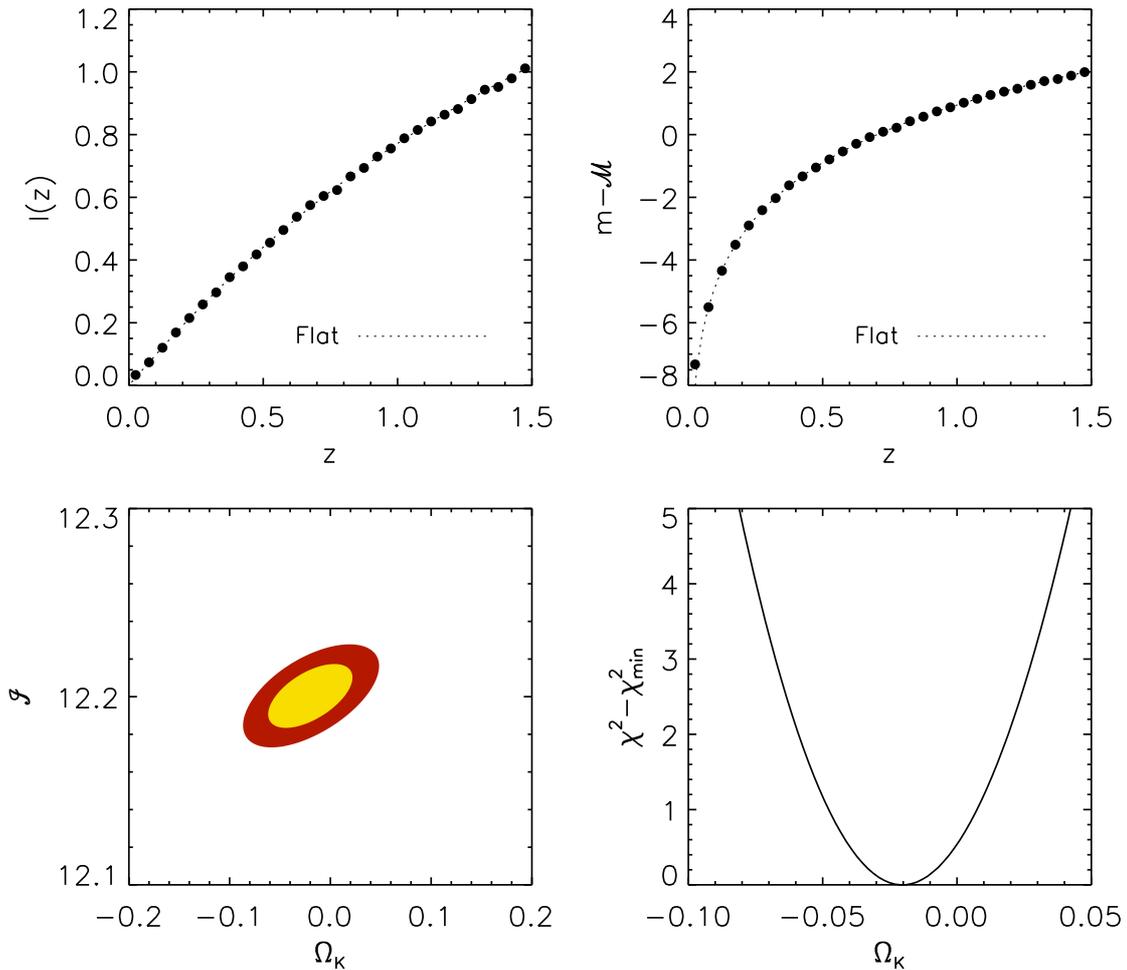}
\caption{\label{fig:omkfit} 
Future constraints on $\ok$ with 
2000 well observed SNe~Ia in the range $ 0.002 \le z \le 1.5$ with
$\sigma_m=0.1$ mag and 2000 galaxy age measurements in the range $ 0.1
\le z \le 1.5$ with a precision of 0.1 Gyr. 
{\em Upper left:} $I(z)$ as derived from galaxy ages.  {\em Upper
right:} $m-\mathcal{M}$ as derived from SN~Ia distances.  {\em Lower left:}
Confidence contours in the $[\ok, \mathcal{I}]$-plane at the 68.3\%
and 95\% CL after marginalisation over $H_0$ assuming
a 1\% Gaussian prior. {\em Lower right:} The resulting $\chi^2-\chi^2_{\rm min}$
giving $\ok =-0.005 \pm 0.054$ (95\% CL. }
\end{center}
\end{figure}

Applying the integral approach to our default future data set, we
obtain the result depicted in figure~\ref{fig:omkfit}.
After marginalisation over $\mathcal{I}$, we find $\ok =-0.005 \pm
0.054$ (95\% CL). Here we have assumed perfect knowledge of
$\mathcal{M}$ and a 1\% prior on $H_0$. In table~\ref{tab:intfut}, we
show constraints at the 95\% CL for different values of $n_{\rm sn}$
and $n_{\rm age}$. From this table we learn two things: First,
increasing only one of the number helps, but is not as effective as
increasing both, i.e., for the numbers used, neither the errors from
SNe or galaxy ages dominate completely the error budget.
Consequently, $n_{\rm sn}=n_{\rm age}$ is
optimal. From an observational point of view it might however be
easier and less expensive to increase one of the two
numbers. Second, for $n_{\rm sn}=n_{\rm age}$, the errors scale as
$n^{-1/2}$. This implies that reaching an error in $\ok$ of $\simeq 0.01$ 
at the 95\% CL requires $n_{\rm sn}=n_{\rm age} \simeq 5\cdot10^4$.

We have also investigated the sensitivity of our results with respect
to the Hubble constant prior in the case where $n_{\rm SN}=n_{\rm
age}=500$. If the value of $H_0$ was perfectly known, we would
anticipate a measurement with an uncertainty of $0.052$ at the 95\%
CL. Assuming a 1\% Gaussian prior increases this number to
0.098. Increasing the prior to 2\% and 5\% results in
$0.110$ and $0.119$, respectively. Our forecast are
clearly not too dependent on our choice of the $H_0$ prior.

The redshift range of the data is important. The higher the redshift,
the more constraining power our method have. Assuming $n_{\rm SN}=n_{\rm
age}=500$ but data distributed in the interval $0.002 \le z
\le 1.0$ we find the uncertainty in $\ok$ to be $0.173$ at the 95\% CL. 
Keeping the number of data points fixed and increasing the maximum
redshift to $2$ and $2.5$ leads to an uncertainty in $\ok$ of $0.065$
and $0.054$ at the 95\% CL, respectively.

%----------------------------------------------------------------------
%----------------------------------------------------------------------

%----------------------------------------------------------------------
\begin{table}[t]
\begin{center}
\caption{Constraints on $\ok$ at the 95\% CL from simulated SN~Ia and 
galaxy age data. The data is uniformly distributed in the redshift
interval $0.002 < z < 1.5$. For SNe~Ia and galaxy ages the error is
$\sigma_m=0.1$ mag and $\sigma_{t_{\rm age}}=0.1$ Gyr,
respectively. The value of the Hubble constant was assumed to be known
to 1\%.}
\label{tab:intfut}
\begin{tabular}{ccccc}
\hline
$n_{\rm SN}$ & $n_{\rm age}=100$ &  $n_{\rm age}=500$ & $n_{\rm age}=1000$ & $n_{\rm age}=2000$ \\      
\hline
100 & 0.188 & 0.188 & 0.165 & 0.156 \\
500 & 0.122 & 0.098 & 0.093 & 0.084 \\
1000 & 0.111 & 0.080 & 0.072 & 0.066 \\
2000 & 0.104 & 0.069 & 0.060 & 0.054 \\
\hline
\end{tabular}
\end{center}
\end{table}

%%%%%%%%%%%%%%%%%%%%%%%%%%%%%%%%%%%%%%%%%%%%%%%%%%%%%%%%%%%%%%%%%%%%%%%
%%%%%%%%%%%%%%%%%%%%%%%%%%%%%%%%%%%%%%%%%%%%%%%%%%%%%%%%%%%%%%%%%%%%%%%
\section{Summary and Discussion }\label{sec:summary}
We have investigated how the large scale spatial curvature of the
Universe can be measured model independently. Cosmological distances
depend on $\ok$ via the expansion history and the geometrical path of
photons.  If the two contributions can be disentangled, $\ok$ can be
measured independently of the energy content of the Universe. We have
pursued two different paths to disentangle the effects of geometry and
expansion.

The \emph{differential} approach, originally suggested by Clarkson,
Cort{\^e}s, and Basset \cite{2007JCAP...08..011C}, allows $\ok$ to be
measured via eq.~(\ref{eq:ok}). In addition to distances, this
approach requires derivatives of distance with respect to redshift and
independent measurements of the expansion history.  The expansion
history can be measured using different probes. Here we focus on the
method proposed by Jimenez and Loeb \cite{jim02}, which utilises
relative galaxy ages.  Cosmological distances with corresponding
redshift derivatives can be derived from SN~Ia observations. 
 
The differential approach requires numerical derivation of data to
obtain $D'(z)$, and in the case of galaxy ages, also $E(z)$. Numerical
derivation of noisy data is far from trivial and we suspect that
errors have often been underestimated in the past. In
\cite{2005PhRvD..71l3001S}, for example, they use a customised method
to obtain $H(z)$ that gives uncomfortably small errors. Basically,
they first group together all galaxies that are within $\Delta z
=0.03$ of each other to get an estimate of the age of the universe at
a given redshift with as many galaxies as possible. The redshift
interval is chosen to be small enough to avoid incorporating galaxies
that have already evolved in age, but large enough to have more than
one galaxy in most of the bins. Age differences are computed only for
those bins in redshift that are separated by more than $z = 0.1$ but
not by more than $z = 0.15$. The first limit is imposed so that the
age evolution between the two bins is larger than the error in the age
determination. The authors claim that this procedure provides a robust
determination of $dz/dt_{\rm age}$. We have not been able to reproduce
their results and have therefore not been able to check the stability
of this claim. It is, however, unlikely that the given uncertainties
in the derived values of $H(z)$ are correct. From a sample of 32
galaxy ages (out of which some are discarded as outliers), you could
form a maximum of 16 pairs of 2 galaxy ages each which could give 8
values of the Hubble parameter. Assuming an error of 10\% in the
galaxy ages, a simple analytical error propagation gives an error of
$\gtrsim 100\%$ in $H(z)$.  A straight line fit in bins give similar
results; at best a 100\% error on $H(z)$ at five different
redshifts. With this (lack of) precision it turns out to be futile to
determine $\ok$.  In \cite{2010JCAP...02..008S}, $dt_{\rm age}/dz$ is
computed by defining the edge in the $t_{\rm age}$ vs $z$-plot formed
by the oldest galaxies in a sample of 24 clusters, yielding two
additional $H(z)$~estimates at $z=0.48$ and $0.9$. We have not checked the
stability of this method but the derived $H(z)$~uncertainties seems to
be more reasonable when compared to the number of galaxy ages
employed.

Galaxy ages as a function of redshift can, alternatively, be used to
measure the integral $I(z)$ via eq.~(\ref{eq:imeas}). Knowledge of
$I(z)$ allows $\ok$ to be measured via the \emph{integral} approach,
see eq.~(\ref{eq:curv5}). This approach requires numerical integration
of noisy data instead of numerical differentiation.  There are
potential problems also with the integral method. The integral approach,
which relies on measurements of lookback time, is not as general as
the differential approach, where, in principle, $E(z)$ can be measured
using several fundamentally different methods. We assume that the
lookback time can be measured via passively evolving galaxies acting
as cosmic chronometers. When converting galaxy ages to lookback times
we make the assumption that the star formation halted at the same
time, i.e., that all passively evolving galaxies share the same
birthday. Even if a population of cosmic chronometers exists, there is
always a risk that our sample is contaminated by other galaxies. Two
unknown quantities enters our calculation of $I(z)$: the constant
offset in time between galaxy age and lookback time and the low
redshift part of the integral over galaxy age. Fortunately, these
unknowns can be handled by a single nuisance parameter
$\mathcal{I}$. The need for an extra nuisance parameter is however
another drawback of the method as it degrades the constraining power
of the data. Another problem with the integral method is that values
of $I(z)$ computed using eq.~(\ref{eq:int}) are correlated. However,
as shown in appendix~\ref{app:err} the correlations decrease with
increasing number of galaxy ages.

Generally, the
differential method is the most effective if we have, in addition to
distance measures, independent $H(z)$  constraints from, e.g., radial
BAO measurements. The integral method, on the other hand, outperforms
the differential method if the expansion history needs to be
reconstructed from lookback time measurements. The reason why this is
not the case for current data is that the derived $H(z)$ from passive
galaxy ages probably have underestimated uncertainties.

The model independent constraints on $\ok$ we present here are rather
weak, $-1 \lesssim \ok \lesssim 1$ at the 95\% CL, and at least for
the differential method, there are reasonable doubts that the
uncertainty in the derived expansion history has been underestimated.
We therefore investigated to what degree a proper analysis of future
data could sharpen current curvature constraints.  Even if the amount
of data is significantly increased, we find that the uncertainty in
$\ok$ will be fairly large, $\simeq 0.1$ at the 95\% CL, at least
compared to (model dependent) CMB constraints. It is therefore
unrealistic that the methods presented in this paper will be able to
constrain, e.g. inflationary models that predict spatial curvature on
the order of $|\ok |\lesssim 10^{-5}$.

In the introduction we noted another proposed method of measuring
$\ok$ model independently using weak gravitational lensing and BAO by
Bernstein \cite{2006ApJ...637..598B}. Forecasts for this method,
assuming a full sky survey of galaxies in the redshift range $0 < z <
3$ supplemented with photometric redshifts, predicts an uncertainty in $\ok$
of $\simeq 0.04$ at the 68.3\% CL. This precision is comparable to what could
be obtained with the integral method with $n_{sn}=1000$ and $n_{\rm age}=500$
(see table~\ref{tab:intfut}).

%%%%%%%%%%%%%%%%%%%%%%%%%%%%%%%%%%%%%%%%%%%%%%%%%%%%%%%%%%%%%%%%%%%%%%%
%%%%%%%%%%%%%%%%%%%%%%%%%%%%%%%%%%%%%%%%%%%%%%%%%%%%%%%%%%%%%%%%%%%%%%%
\acknowledgments
EM acknowledge support for this study by the Swedish Research
Council. We thank Licia Verde for kindly providing us with galaxy ages
in electronic form. JJ would like to thank Rahman Amanullah for fruitful 
discussions. 

%%%%%%%%%%%%%%%%%%%%%%%%%%%%%%%%%%%%%%%%%%%%%%%%%%%%%%%%%%%%%%%%%%%%%%%
%%%%%%%%%%%%%%%%%%%%%%%%%%%%%%%%%%%%%%%%%%%%%%%%%%%%%%%%%%%%%%%%%%%%%%%
\appendix
%%%%%%%%%%%%%%%%%%%%%%%%%%%%%%%%%%%%%%%%%%%%%%%%%%%%%%%%%%%%%%%%%%%%%%%
%%%%%%%%%%%%%%%%%%%%%%%%%%%%%%%%%%%%%%%%%%%%%%%%%%%%%%%%%%%%%%%%%%%%%%%
\section{Lookback time and galaxy ages \label{app:lookb}}
The integral $I(z)$ can, according to eq.~(\ref{eq:iage}), be obtained
from measurements of lookback time as function of redshift. We assume
that the lookback time can be measured via the age of galaxies, or
rather the age of the stellar population in galaxies.

To understand the relationship between lookback time and galaxy age,
we write the present age of the Universe as
\begin{eqnarray}
t_0-t_\infty & = & -\int_0^\infty \frac{dt}{dz}dz \\
& = & (t_0-t_z) + (t_z -t_{\rm c})+(t_{\rm c}-t_\infty),
\end{eqnarray}
where $t_0-t_z$ is the lookback time, $t_z-t_{\rm c}$ is the age of a
galaxy at redshift $z$, which was created at redshift $z_{\rm c}$, and
$t_{\rm c}-t_\infty$ is the age of the Universe when the passively
evolving stellar population of the galaxy was created.

We naively assume that all the galaxies share a common time of star
formation. In that case
\begin{equation}
(t_0-t_z)=T-(t_z-t_{\rm c}),
\label{eq:const}
\end{equation}
where the constant $T$ is the age of the Universe at redshift $z_{\rm
c}$ subtracted from the present age of the Universe. The lookback time
and the galaxy age are hence related by an additive
constant. Consequently galaxy ages are related to $\tau$ through
\begin{equation}
\tau=H_0(T-t_{\rm age}).
\label{eq:iobs}
\end{equation}

\section{Derivation of error in the integral approximation \label{app:err}}
We use eq.~(\ref{eq:int}) to compute the area under the graph of
$t_{\rm age}(z)$. Let us make some simplifying assumptions that will
allow us to draw some conclusions about the accuracy of the
approximation. We assume all the measurements $t_{{\rm age},i}$ at redshifts $z_i$
to be separated by $\Delta z$. Furthermore we assume the error
$\sigma_{t_{\rm age}}$ to be the same for all $t_{{\rm age},i}$. 
Values of $F(z)$ computed at
different redshifts using eq.~(\ref{eq:int}) are obviously
correlated. With the previous assumptions, we can derive an
approximate formula for the covariance matrix using error
propagation,
\begin{equation}
U_{jk} \simeq \sum_{i=0}^{{\rm min}(j,k)} \frac{\partial F_j}
{\partial t_{{\rm age},i}} 
\frac{\partial F_k}{\partial t_{{\rm age},i}} 
\sigma_{t_{\rm age}}^2 =
\left[{\rm min}(j,k)-\frac{1}{4}(1+\delta_{jk}) \right]\Delta z^2 
\sigma_{t_{\rm age}}^2 \lesssim 
{\rm min}(j,k)\Delta z^2 \sigma_{t_{\rm age}}^2.
\end{equation}  
The assumption that the points are evenly distributed in redshift
means that $dn/dz=\alpha$ is a constant. Consequently  $k=\alpha(z_k-z_{\rm
min})$ and $\Delta z = \alpha^{-1}$. We can therefore rewrite the
covariance in terms of redshift as
\begin{equation}
U_{jk} \lesssim \frac{\sigma_{t_{\rm age}}^2}{\alpha}\left[{\rm
min}(z_j,z_k)-z_{\rm min} \right].
\label{eq:cov}
\end{equation}
At redshift $z_k$ the variance is hence 
\begin{equation}
\sigma_F^2 \equiv U_{kk}\lesssim \frac{\sigma_{t_{\rm age}}^2}{\alpha}\left(z_k-z_{\rm min}\right).
\label{eq:var}
\end{equation}
We note that $U_{jk} \lesssim U_{kk}$.

%%%%%%%%%%%%%%%%%%%%%%%%%%%%%%%%%%%%%%%%%%%%%%%%%%%%%%%%%%%%%%%%%%%%%%%
%%%%%%%%%%%%%%%%%%%%%%%%%%%%%%%%%%%%%%%%%%%%%%%%%%%%%%%%%%%%%%%%%%%%%%%
\bibliographystyle{JHEP}
\bibliography{paper}

\providecommand{\href}[2]{#2}\begingroup\raggedright\begin{thebibliography}{10}

\bibitem{2009ApJS..180..330K}
E.~{Komatsu}, J.~{Dunkley}, M.~R. {Nolta}, C.~L. {Bennett}, B.~{Gold},
  G.~{Hinshaw}, N.~{Jarosik}, D.~{Larson}, M.~{Limon}, L.~{Page}, D.~N.
  {Spergel}, M.~{Halpern}, R.~S. {Hill}, A.~{Kogut}, S.~S. {Meyer}, G.~S.
  {Tucker}, J.~L. {Weiland}, E.~{Wollack}, and E.~L. {Wright}, {\it {Five-Year
  Wilkinson Microwave Anisotropy Probe Observations: Cosmological
  Interpretation}},  {\em \apjs} {\bf 180} (2009) 330,
  [\href{http://xxx.lanl.gov/abs/0803.0547}{{\tt arXiv:0803.0547}}].

\bibitem{2009arXiv0908.4274K}
R.~{Kessler}, A.~C. {Becker}, D.~{Cinabro}, J.~{Vanderplas}, J.~A. {Frieman},
  J.~{Marriner}, T.~M. {Davis}, B.~{Dilday}, J.~{Holtzman}, S.~W. {Jha},
  H.~{Lampeitl}, M.~{Sako}, M.~{Smith}, C.~{Zheng}, R.~C. {Nichol},
  B.~{Bassett}, R.~{Bender}, D.~L. {Depoy}, M.~{Doi}, E.~{Elson}, A.~V.
  {Filippenko}, R.~J. {Foley}, P.~M. {Garnavich}, U.~{Hopp}, Y.~{Ihara},
  W.~{Ketzeback}, W.~{Kollatschny}, K.~{Konishi}, J.~L. {Marshall}, R.~J.
  {McMillan}, G.~{Miknaitis}, T.~{Morokuma}, E.~{M{\"o}rtsell}, K.~{Pan}, J.~L.
  {Prieto}, M.~W. {Richmond}, A.~G. {Riess}, R.~{Romani}, D.~P. {Schneider},
  J.~{Sollerman}, N.~{Takanashi}, K.~{Tokita}, K.~{van der Heyden}, J.~C.
  {Wheeler}, N.~{Yasuda}, and D.~{York}, {\it {First-Year Sloan Digital Sky
  Survey-II Supernova Results: Hubble Diagram and Cosmological Parameters}},
  {\em \apjs} {\bf 185} (2009) 32,
  [\href{http://xxx.lanl.gov/abs/0908.4274}{{\tt arXiv:0908.4274}}].

\bibitem{2006PhRvD..73b3503K}
L.~{Knox}, {\it {Precision measurement of the mean curvature}},  {\em \prd}
  {\bf 73} (2006), no.~2 023503.

\bibitem{cro03}
J.~L. {Crooks}, J.~O. {Dunn}, P.~H. {Frampton}, H.~R. {Norton}, and
  T.~{Takahashi}, {\it {Cosmic degeneracy with dark energy equation of state}},
   {\em \astropart} {\bf 20} (2003) 361.

\bibitem{pol05}
D.~{Polarski} and A.~{Ranquet}, {\it {On the equation of state of dark
  energy}},  {\em \physlettb} {\bf 627} (2005) 1.

\bibitem{2007JCAP...08..011C}
C.~{Clarkson}, M.~{Cort{\^e}s}, and B.~{Bassett}, {\it {Dynamical dark energy
  or simply cosmic curvature?}},  {\em Journal of Cosmology and Astro-Particle
  Physics} {\bf 8} (2007) 11.

\bibitem{hua07}
Z.~{Huang}, B.~{Wang}, and R.~{Su}, {\it {Uncertainty on Determining the Dark
  Energy Equation of State due to the Spatial Curvature}},  {\em \intmoda} {\bf
  22} (2007) 1819.

\bibitem{2009arXiv0910.0252B}
G.~{Barenboim}, E.~{Fern{\'a}ndez Mart{\'{\i}}nez}, O.~{Mena}, and L.~{Verde},
  {\it {The dark side of curvature}},  {\em \jcap} {\bf 3} (2010) 8,
  [\href{http://xxx.lanl.gov/abs/0910.0252}{{\tt arXiv:0910.0252}}].

\bibitem{gon05}
Y.~{Gong} and Y.~{Zhang}, {\it {Probing the curvature and dark energy}},  {\em
  \physrevd} {\bf 72} (2005) 043518.

\bibitem{ich06a}
K.~{Ichikawa} and T.~{Takahashi}, {\it {Dark energy evolution and the curvature
  of the universe from recent observations}},  {\em \physrevd} {\bf 73} (2006)
  083526.

\bibitem{ich06b}
K.~{Ichikawa}, M.~{Kawasaki}, T.~{Sekiguchi}, and T.~{Takahashi}, {\it
  {Implications of dark energy parametrizations for the determination of the
  curvature of the universe}},  {\em \jcap} {\bf 12} (2006) 5.

\bibitem{wri07}
E.~L. {Wright}, {\it {Constraints on Dark Energy from Supernovae, Gamma-Ray
  Bursts, Acoustic Oscillations, Nucleosynthesis, Large-Scale Structure, and
  the Hubble Constant}},  {\em \astrophys} {\bf 664} (2007) 633.

\bibitem{zha07}
G.~{Zhao}, J.~{Xia}, H.~{Li}, C.~{Tao}, J.~{Virey}, Z.~{Zhu}, and X.~{Zhang},
  {\it {Probing for dynamics of dark energy and curvature of universe with
  latest cosmological observations}},  {\em \physlettb} {\bf 648} (2007) 8.

\bibitem{rei09}
B.~A. {Reid}, W.~J. {Percival}, D.~J. {Eisenstein}, L.~{Verde}, D.~N.
  {Spergel}, R.~A. {Skibba}, N.~A. {Bahcall}, T.~{Budavari}, J.~A. {Frieman},
  M.~{Fukugita}, J.~R. {Gott}, J.~E. {Gunn}, {\v Z}.~{Ivezi{\'c}}, G.~R.
  {Knapp}, R.~G. {Kron}, R.~H. {Lupton}, T.~A. {McKay}, A.~{Meiksin}, R.~C.
  {Nichol}, A.~C. {Pope}, D.~J. {Schlegel}, D.~P. {Schneider}, C.~{Stoughton},
  M.~A. {Strauss}, A.~S. {Szalay}, M.~{Tegmark}, M.~S. {Vogeley}, D.~H.
  {Weinberg}, D.~G. {York}, and I.~{Zehavi}, {\it {Cosmological constraints
  from the clustering of the Sloan Digital Sky Survey DR7 luminous red
  galaxies}},  {\em \mnras} {\bf 404} (2010) 60,
  [\href{http://xxx.lanl.gov/abs/0907.1659}{{\tt arXiv:0907.1659}}].

\bibitem{vik09}
A.~{Vikhlinin}, A.~V. {Kravtsov}, R.~A. {Burenin}, H.~{Ebeling}, W.~R.
  {Forman}, A.~{Hornstrup}, C.~{Jones}, S.~S. {Murray}, D.~{Nagai},
  H.~{Quintana}, and A.~{Voevodkin}, {\it {Chandra Cluster Cosmology Project
  III: Cosmological Parameter Constraints}},  {\em \astrophys} {\bf 692} (2009)
  1060, [\href{http://xxx.lanl.gov/abs/0812.2720}{{\tt arXiv:0812.2720}}].

\bibitem{2010arXiv1001.4538K}
E.~{Komatsu}, K.~M. {Smith}, J.~{Dunkley}, C.~L. {Bennett}, B.~{Gold},
  G.~{Hinshaw}, N.~{Jarosik}, D.~{Larson}, M.~R. {Nolta}, L.~{Page}, D.~N.
  {Spergel}, M.~{Halpern}, R.~S. {Hill}, A.~{Kogut}, M.~{Limon}, S.~S. {Meyer},
  N.~{Odegard}, G.~S. {Tucker}, J.~L. {Weiland}, E.~{Wollack}, and E.~L.
  {Wright}, {\it {Seven-year Wilkinson Microwave Anisotropy Probe (WMAP)
  Observations: Cosmological Interpretation}},  {\em \apjs} {\bf 192} (2011)
  18, [\href{http://xxx.lanl.gov/abs/1001.4538}{{\tt arXiv:1001.4538}}].

\bibitem{ich07}
K.~{Ichikawa} and T.~{Takahashi}, {\it {Dark energy parametrizations and the
  curvature of the universe}},  {\em \jcap} {\bf 2} (2007) 1.

\bibitem{2009JCAP...01..044M}
E.~{M{\"o}rtsell} and C.~{Clarkson}, {\it {Model independent constraints on the
  cosmological expansion rate}},  {\em Journal of Cosmology and Astro-Particle
  Physics} {\bf 1} (2009) 44, [\href{http://xxx.lanl.gov/abs/0811.0981}{{\tt
  arXiv:0811.0981}}].

\bibitem{mor09}
M.~J. {Mortonson}, {\it {Testing flatness of the universe with probes of cosmic
  distances and growth}},  {\em \physrevd} {\bf 80} (2009) 123504,
  [\href{http://xxx.lanl.gov/abs/0908.0346}{{\tt arXiv:0908.0346}}].

\bibitem{2006ApJ...637..598B}
G.~{Bernstein}, {\it {Metric Tests for Curvature from Weak Lensing and Baryon
  Acoustic Oscillations}},  {\em \apj} {\bf 637} (2006) 598.

\bibitem{2009ApJ...703.1374S}
J.~{Sollerman}, E.~{M{\"o}rtsell}, T.~M. {Davis}, M.~{Blomqvist}, B.~{Bassett},
  A.~C. {Becker}, D.~{Cinabro}, A.~V. {Filippenko}, R.~J. {Foley},
  J.~{Frieman}, P.~{Garnavich}, H.~{Lampeitl}, J.~{Marriner}, R.~{Miquel},
  R.~C. {Nichol}, M.~W. {Richmond}, M.~{Sako}, D.~P. {Schneider}, M.~{Smith},
  J.~T. {Vanderplas}, and J.~C. {Wheeler}, {\it {First-Year Sloan Digital Sky
  Survey-II (SDSS-II) Supernova Results: Constraints on Nonstandard
  Cosmological Models}},  {\em \apj} {\bf 703} (2009) 1374,
  [\href{http://xxx.lanl.gov/abs/0908.4276}{{\tt arXiv:0908.4276}}].

\bibitem{2009arXiv0909.4723B}
M.~{Blomqvist} and E.~{M{\"o}rtsell}, {\it {Supernovae as seen by off-center
  observers in a local void}},  {\em \jcap} {\bf 5} (2010) 6,
  [\href{http://xxx.lanl.gov/abs/0909.4723}{{\tt arXiv:0909.4723}}].

\bibitem{2008JCAP...06..027B}
M.~{Blomqvist}, E.~{M{\"o}rtsell}, and S.~{Nobili}, {\it {Probing dark energy
  inhomogeneities with supernovae}},  {\em Journal of Cosmology and
  Astro-Particle Physics} {\bf 6} (2008) 27,
  [\href{http://xxx.lanl.gov/abs/0806.0496}{{\tt arXiv:0806.0496}}].

\bibitem{cla08}
C.~{Clarkson}, B.~{Bassett}, and T.~{Lu}, {\it {A General Test of the
  Copernican Principle}},  {\em \physrevl} {\bf 101} (2008) 011301,
  [\href{http://xxx.lanl.gov/abs/0712.3457}{{\tt arXiv:0712.3457}}].

\bibitem{sha09}
A.~{Shafieloo} and C.~{Clarkson}, {\it {Model independent tests of the standard
  cosmological model}},  {\em \prd} {\bf 81} (2010), no.~8 083537,
  [\href{http://xxx.lanl.gov/abs/0911.4858}{{\tt arXiv:0911.4858}}].

\bibitem{jim02}
R.~{Jimenez} and A.~{Loeb}, {\it {Constraining Cosmological Parameters Based on
  Relative Galaxy Ages}},  {\em \astrophys} {\bf 573} (2002) 37.

\bibitem{2005PhRvD..71j3513W}
Y.~{Wang} and M.~{Tegmark}, {\it {Uncorrelated measurements of the cosmic
  expansion history and dark energy from supernovae}},  {\em \prd} {\bf 71}
  (2005), no.~10 103513.

\bibitem{2007ApJ...659...98R}
A.~G. {Riess}, L.~{Strolger}, S.~{Casertano}, H.~C. {Ferguson}, B.~{Mobasher},
  B.~{Gold}, P.~J. {Challis}, A.~V. {Filippenko}, S.~{Jha}, W.~{Li},
  J.~{Tonry}, R.~{Foley}, R.~P. {Kirshner}, M.~{Dickinson}, E.~{MacDonald},
  D.~{Eisenstein}, M.~{Livio}, J.~{Younger}, C.~{Xu}, T.~{Dahl{\'e}n}, and
  D.~{Stern}, {\it {New Hubble Space Telescope Discoveries of Type Ia
  Supernovae at $z \ge 1$: Narrowing Constraints on the Early Behavior of Dark
  Energy}},  {\em \apj} {\bf 659} (2007) 98.

\bibitem{2010ApJ...716..712A}
R.~{Amanullah}, C.~{Lidman}, D.~{Rubin}, G.~{Aldering}, P.~{Astier},
  K.~{Barbary}, M.~S. {Burns}, A.~{Conley}, K.~S. {Dawson}, S.~E. {Deustua},
  M.~{Doi}, S.~{Fabbro}, L.~{Faccioli}, H.~K. {Fakhouri}, G.~{Folatelli}, A.~S.
  {Fruchter}, H.~{Furusawa}, G.~{Garavini}, G.~{Goldhaber}, A.~{Goobar}, D.~E.
  {Groom}, I.~{Hook}, D.~A. {Howell}, N.~{Kashikawa}, A.~G. {Kim}, R.~A.
  {Knop}, M.~{Kowalski}, E.~{Linder}, J.~{Meyers}, T.~{Morokuma}, S.~{Nobili},
  J.~{Nordin}, P.~E. {Nugent}, L.~{{\"O}stman}, R.~{Pain}, N.~{Panagia},
  S.~{Perlmutter}, J.~{Raux}, P.~{Ruiz-Lapuente}, A.~L. {Spadafora},
  M.~{Strovink}, N.~{Suzuki}, L.~{Wang}, W.~M. {Wood-Vasey}, N.~{Yasuda}, and
  T.~{Supernova Cosmology Project}, {\it {Spectra and Hubble Space Telescope
  Light Curves of Six Type Ia Supernovae at 0.511 $\lt$ z $\lt$ 1.12 and the
  Union2 Compilation}},  {\em \apj} {\bf 716} (2010) 712,
  [\href{http://xxx.lanl.gov/abs/1004.1711}{{\tt arXiv:1004.1711}}].

\bibitem{2008arXiv0804.4142K}
M.~{Kowalski}, D.~{Rubin}, G.~{Aldering}, R.~J. {Agostinho}, A.~{Amadon},
  R.~{Amanullah}, C.~{Balland}, K.~{Barbary}, G.~{Blanc}, P.~J. {Challis},
  A.~{Conley}, N.~V. {Connolly}, R.~{Covarrubias}, K.~S. {Dawson}, S.~E.
  {Deustua}, R.~{Ellis}, S.~{Fabbro}, V.~{Fadeyev}, X.~{Fan}, B.~{Farris},
  G.~{Folatelli}, B.~L. {Frye}, G.~{Garavini}, E.~L. {Gates}, L.~{Germany},
  G.~{Goldhaber}, B.~{Goldman}, A.~{Goobar}, D.~E. {Groom}, J.~{Haissinski},
  D.~{Hardin}, I.~{Hook}, S.~{Kent}, A.~G. {Kim}, R.~A. {Knop}, C.~{Lidman},
  E.~V. {Linder}, J.~{Mendez}, J.~{Meyers}, G.~J. {Miller}, M.~{Moniez}, A.~M.
  {Mourao}, H.~{Newberg}, S.~{Nobili}, P.~E. {Nugent}, R.~{Pain},
  O.~{Perdereau}, S.~{Perlmutter}, M.~M. {Phillips}, V.~{Prasad}, R.~{Quimby},
  N.~{Regnault}, J.~{Rich}, E.~P. {Rubenstein}, P.~{Ruiz-Lapuente}, F.~D.
  {Santos}, B.~E. {Schaefer}, R.~A. {Schommer}, R.~C. {Smith}, A.~M.
  {Soderberg}, A.~L. {Spadafora}, L.~. {Strolger}, M.~{Strovink}, N.~B.
  {Suntzeff}, N.~{Suzuki}, R.~C. {Thomas}, N.~A. {Walton}, L.~{Wang}, W.~M.
  {Wood-Vasey}, and J.~L. {Yun}, {\it {Improved Cosmological Constraints from
  New, Old and Combined Supernova Datasets}},  {\em ArXiv e-prints} {\bf 804}
  (2008) [\href{http://xxx.lanl.gov/abs/0804.4142}{{\tt arXiv:0804.4142}}].

\bibitem{dun96}
J.~{Dunlop}, J.~{Peacock}, H.~{Spinrad}, A.~{Dey}, R.~{Jimenez}, D.~{Stern},
  and R.~{Windhorst}, {\it {A 3.5-Gyr-old galaxy at redshift 1.55}},  {\em
  \nat} {\bf 381} (1996) 581.

\bibitem{spi97}
H.~{Spinrad}, A.~{Dey}, D.~{Stern}, J.~{Dunlop}, J.~{Peacock}, R.~{Jimenez},
  and R.~{Windhorst}, {\it {LBDS 53W091: an Old, Red Galaxy at z=1.552}},  {\em
  \astrophys} {\bf 484} (1997) 581.

\bibitem{cow99}
L.~L. {Cowie}, A.~{Songaila}, and A.~J. {Barger}, {\it {Evidence for a Gradual
  Decline in the Universal Rest-Frame Ultraviolet Luminosity Density for Z $<$
  1}},  {\em \astronj} {\bf 118} (1999) 603.

\bibitem{hea04}
A.~{Heavens}, B.~{Panter}, R.~{Jimenez}, and J.~{Dunlop}, {\it {The
  star-formation history of the Universe from the stellar populations of nearby
  galaxies}},  {\em \nat} {\bf 428} (2004) 625.

\bibitem{tho05}
D.~{Thomas}, C.~{Maraston}, R.~{Bender}, and C.~{Mendes de Oliveira}, {\it {The
  Epochs of Early-Type Galaxy Formation as a Function of Environment}},  {\em
  \astrophys} {\bf 621} (2005) 673.

\bibitem{tre05}
T.~{Treu}, R.~S. {Ellis}, T.~X. {Liao}, P.~G. {van Dokkum}, P.~{Tozzi},
  A.~{Coil}, J.~{Newman}, M.~C. {Cooper}, and M.~{Davis}, {\it {The Assembly
  History of Field Spheroidals: Evolution of Mass-to-Light Ratios and
  Signatures of Recent Star Formation}},  {\em \astrophys} {\bf 633} (2005)
  174.

\bibitem{pan07}
B.~{Panter}, R.~{Jimenez}, A.~F. {Heavens}, and S.~{Charlot}, {\it {The star
  formation histories of galaxies in the Sloan Digital Sky Survey}},  {\em
  \mnras} {\bf 378} (2007) 1550.

\bibitem{sto95}
A.~{Stockton}, M.~{Kellogg}, and S.~E. {Ridgway}, {\it {The nature of the
  stellar continuum in the radio galaxy 3C 65}},  {\em \apjl} {\bf 443} (1995)
  L69.

\bibitem{sto01}
A.~{Stockton}, {\it {The Oldest Stellar Populations at z $\sim$ 1.5}},  in {\em
  Astrophysical Ages and Times Scales} ({T.~von Hippel, C.~Simpson, \&
  N.~Manset}, ed.), vol.~245 of {\em \aspc}, p.~517, 2001.

\bibitem{alc01}
J.~S. {Alcaniz} and J.~A.~S. {Lima}, {\it {Dark Energy and the Epoch of Galaxy
  Formation}},  {\em \apjl} {\bf 550} (2001) L133.

\bibitem{jim03}
R.~{Jimenez}, L.~{Verde}, T.~{Treu}, and D.~{Stern}, {\it {Constraints on the
  Equation of State of Dark Energy and the Hubble Constant from Stellar Ages
  and the Cosmic Microwave Background}},  {\em \astrophys} {\bf 593} (2003)
  622.

\bibitem{2005PhRvD..71l3001S}
J.~{Simon}, L.~{Verde}, and R.~{Jimenez}, {\it {Constraints on the redshift
  dependence of the dark energy potential}},  {\em \prd} {\bf 71} (2005),
  no.~12 123001.

\bibitem{2010JCAP...02..008S}
D.~{Stern}, R.~{Jimenez}, L.~{Verde}, M.~{Kamionkowski}, and S.~A. {Stanford},
  {\it {Cosmic chronometers: constraining the equation of state of dark energy.
  I: H(z) measurements}},  {\em Journal of Cosmology and Astro-Particle
  Physics} {\bf 2} (2010) 8, [\href{http://xxx.lanl.gov/abs/0907.3149}{{\tt
  arXiv:0907.3149}}].

\bibitem{san62}
A.~{Sandage}, {\it {The Change of Redshift and Apparent Luminosity of Galaxies
  due to the Deceleration of Selected Expanding Universes.}},  {\em \astrophys}
  {\bf 136} (1962) 319.

\bibitem{bon06}
C.~{Bonvin}, R.~{Durrer}, and M.~{Kunz}, {\it {Dipole of the Luminosity
  Distance: A Direct Measure of H(z)}},  {\em \physrevl} {\bf 96} (2006)
  191302.

\bibitem{2009MNRAS.399.1663G}
E.~{Gazta{\~n}aga}, A.~{Cabr{\'e}}, and L.~{Hui}, {\it {Clustering of luminous
  red galaxies - IV. Baryon acoustic peak in the line-of-sight direction and a
  direct measurement of H(z)}},  {\em \mnras} {\bf 399} (2009) 1663,
  [\href{http://xxx.lanl.gov/abs/0807.3551}{{\tt arXiv:0807.3551}}].

\bibitem{2009arXiv0901.1219M}
J.~{Miralda-Escude}, {\it {Comment on the claimed radial BAO detection by
  Gaztanaga et al}},  {\em ArXiv e-prints} (2009)
  [\href{http://xxx.lanl.gov/abs/0901.1219}{{\tt arXiv:0901.1219}}].

\bibitem{2010arXiv1011.2729C}
A.~{Cabre} and E.~{Gazta{\~n}aga}, {\it {Have Baryonic Acoustic Oscillations in
  the galaxy distribution really been measured?}},  {\em ArXiv e-prints} (2010)
  [\href{http://xxx.lanl.gov/abs/1011.2729}{{\tt arXiv:1011.2729}}].

\bibitem{rie09}
A.~G. {Riess}, L.~{Macri}, S.~{Casertano}, M.~{Sosey}, H.~{Lampeitl}, H.~C.
  {Ferguson}, A.~V. {Filippenko}, S.~W. {Jha}, W.~{Li}, R.~{Chornock}, and
  D.~{Sarkar}, {\it {A Redetermination of the Hubble Constant with the Hubble
  Space Telescope from a Differential Distance Ladder}},  {\em \astrophys} {\bf
  699} (2009) 539, [\href{http://xxx.lanl.gov/abs/0905.0695}{{\tt
  arXiv:0905.0695}}].

\bibitem{2005ApJ...631....1G}
K.~{Glazebrook} and C.~{Blake}, {\it {Measuring the Cosmic Evolution of Dark
  Energy with Baryonic Oscillations in the Galaxy Power Spectrum}},  {\em \apj}
  {\bf 631} (2005) 1.

\end{thebibliography}\endgroup

\end{document}